\documentclass[fleqn,usenatbib]{mnras}
\usepackage[T1]{fontenc}
\usepackage{ae,aecompl}
\usepackage{subfigure}
\usepackage{graphicx}
\usepackage{amsmath}	
\usepackage{amssymb}	
\usepackage{tablefootnote}
\usepackage{bm}

\newcommand{\Kelvin}{\:\mathrm{K}} 
\newcommand{\kms}{\:\mathrm{km}\:\mathrm{sec}^{-1}} 
\newcommand{\nuAng}{\textup{\AA}}
\newcommand{\Ang}{\:\textup{\AA}}
\newcommand{\Msun}{$\;\mathrm{M}_{\sun}\:$} 
\newcommand{\Msunp}{$\;\mathrm{M}_{\sun}$} 
\newcommand{\Rsun}{$\;\mathrm{R}_{\sun}\:$} 
\newcommand{\Rsunp}{$\;\mathrm{R}_{\sun}$} 
\newcommand{\textsoft}[1]{\textsc{#1}} 

\date{\today}

\title[Shell HVF in SN~2011fe]{A Compact Circumstellar Shell as the Source of High-velocity Features in SN~2011fe}

\author[B. W. Mulligan \& J. C. Wheeler]{
Brian W. Mulligan$^{1}$\thanks{E-mail: bwmulligan@astro.as.utexas.edu} \&
J. Craig Wheeler$^{1}$\thanks{E-mail: wheel@astro.as.utexas.edu}\\
$^{1}$Department of Astronomy, University of Texas at Austin, Austin, TX 78712, USA
}

\date{Accepted 2017 November 29. Received 2017 November 28; in original form 2017 May 14.}
\pubyear{2017}

\begin{document}

\label{firstpage}
\pagerange{\pageref{firstpage}-\pageref{lastpage}}
\maketitle

\begin{abstract}
High-velocity features (HVF), especially of \ion{Ca}{II}, are frequently seen in Type~Ia supernova observed prior to B-band maximum (Bmax). These HVF evolve in velocity from more than $25,000\kms$, in the days after first light, to about $18,000\kms$ near Bmax. To recreate the evolution of the \ion{Ca}{II} near-infrared triplet (CaNIR) HVF in SN~2011fe, we consider the interaction between a model Type~Ia supernova and compact circumstellar shells with masses between 0.003\Msun and 0.012\Msunp. We fit the observed CaNIR feature using synthetic spectra generated from the models using \textsoft{syn++}. The CaNIR feature is better explained by the supernova model interacting with a shell than the model without a shell, with a shell of mass 0.005\Msun tending to be better fitting than the other shells. The evolution of the optical depth of CaNIR suggests that the ionization state of calcium within the ejecta and shell is not constant. We discuss the method used to measure the observed velocity of CaNIR and other features and conclude that HVF or other components can be falsely identified. We briefly discuss the possible origin of the shells and the implications for the progenitor system of the supernova.
\end{abstract}



\section{Introduction}
Type Ia supernovae (SN~Ia) represent an extreme end point of binary stellar evolution and a challenge to thermonuclear combustion physics. Their spectral and photometric characteristics are sufficiently uniform to allow them to be used as calibrated candles. Further study of SN~Ia is important for a deeper understanding of binary stellar evolution, thermonuclear explosions, and to buttress the use of SN~Ia as cosmological probes. A SN~Ia progenitor has yet to be detected prior to the explosion, therefore it is necessary to use the information from the explosion itself to understand these systems. Photometry and spectroscopy of the system within the first days and weeks after the explosion are especially important, as the outermost layers have the strongest effect during this phase. SN~2011fe was fortuitously discovered within about a day after the explosion \citep{Nugent2011}, and is among the nearest to the time of the explosion that a SN~Ia has been observed.

In SN~Ia observed before peak brightness, absorption features of calcium, silicon, and other elements are observed with two components: a high-velocity component with a velocity between $18,000$--$35,000\kms$, referred to as a high-velocity feature (HVF), and a lower-velocity component with a velocity consistently about $7,000\kms$ slower than the HVF \citep{1999ApJ...525..881H, 2003ApJ...591.1110W, 2004ApJ...601.1019T, 2005ApJ...623L..37M, 2006ApJ...636..400Q, 2012ApJ...752L..26P, Marion2013, 2013ApJ...770...29C, 2014MNRAS.437..338C, 2014MNRAS.444.3258M, 2015MNRAS.451.1973S, 2015ApJS..220...20Z, 2016ApJ...826..211Z}. The lower-velocity component, typically referred to as the photospheric-velocity feature (PVF) is the strongest component at and after peak brightness. Both components slow over time, with the HVF reaching an asymptote of $17$--$23,000\kms$ and the PVF reaching $8$--$15,000\kms$ at about the time of peak brightness. The HVF is strongest, in terms of the pseudo-equivalent width (pEW), at the earliest epochs, fading away entirely by about one week after peak brightness. The PVF remains relatively constant in strength over all epochs over which the HVF is visible \citep{2014MNRAS.437..338C, 2015MNRAS.451.1973S}.

The PVF is associated with either unburned carbon and oxygen or material that is synthesized during the explosion, and is reasonably well explained by existing explosion models. The HVF is distinct from the PVF and is a result of material that is physically separate from the material that causes the PVF. The HVF also exhibit a polarization of 0.1 -- 1 per cent \citep{2003ApJ...591.1110W, 2008ARA&A..46..433W, 2009A&A...508..229P}, implying that the material causing the HVF is clumpy or otherwise assymetric. The composition, structure, and source of the material that causes the HVF is unclear. Any model that explains these features must explain the polarization and the temporal evolution of the appearance of both the HVF and PVF. Explanatory models fall into three broad categories: material that is part of the `normal' ejecta but has stronger absorption features due to enhanced populations of the ion- and atomic- state(s) responsible for the feature, material ejected at high velocity during the explosion, and material swept up by the ejecta after the explosion.

The first category, enhanced populations at high velocity, relies on a solar or near-solar abundance of metals in the outer layers that are not in local thermodynamic equilibrium. In this category of models there must be an enhancement of the population of the species responsible for the HVF compared to similarly enriched material at slightly lower velocities. The evolution in velocity of the HVF in these models is due to the evolution in position of the population enhancement. The underlying cause of this shift is entirely dependent on the details of the physics of the excitation and ionization of the material. One challenge of this class of models is that there is no natural explanation for the population enhancements to be clumpy, thereby failing to explain the polarization of the HVF.

One set of delayed-detonation explosion models evaluated by \citet{2013MNRAS.429.2127B} show that the models with the lowest density at the time of transition from deflagration to detonation will have a high-velocity \ion{Ca}{II} near-infrared triplet (CaNIR) feature, though they predict a stronger than observed HVF component. These models were evaluated only near the time of bolometric maximum, so the temporal evolution and appearance at early epochs of the CaNIR feature for these models is not yet known.

The second category includes bullets of material that are ejected at high velocity during the detonation of the carbon and oxygen in the progenitor white dwarf. Such bullets could contain enhanced quantities of newly synthesized calcium, silicon, or other elements and can potentially explain the polarization of the HVF. This model would result in HVF at an early epoch that likely fade over time as the material expands. Testing this category of models requires hydrodynamic simulations of the explosion with sufficient resolution to capture any material ejected in this way. This category of models has not yet been evaluated.

The final set of models include interaction with a wind or an essentially static circumstellar medium. These models explain the HVF as the result of the density enhancement at the contact discontinuity, the clumpiness by Rayleigh-Taylor instability at the contact discontinuity, and velocity evolution by the slowing of the contact discontinuity as more material is swept up, the changing density of the material at the contact discontinuity as it expands, and evolving ion- and atomic- state populations in the material.

\citet{Gerardy2004} considered a simulation of the interaction between the ejecta and a wind, demonstrating that models with a total mass of swept up material of about $2$--$5\times 10^{-2}$\Msun do show evidence of high-velocity features near the time of maximum light. The temporal evolution of the velocity and strength of the HVF in these models has not been evaluated. Models such as this involving an ongoing interaction with a wind or extended shell are now disfavored due to predicted radio emission \citep{2016ApJ...823..100H} during the interaction that has not been observed in SN~2011fe\citep{2012ApJ...750..164C}.

Interaction with a radially compact shell allows for high-velocity material that does not produce excess radiation for more than an hour or two after the explosion. To detect the radiation produced by such an interaction would require very early observation of the supernova. \citet{2012ApJ...744L..17B} place constraints on the \textit{g}-band flux of SN~2011fe at 3.92~h after the explosion, with the time of explosion extrapolated from the rising light curve. This constraint could be relaxed by a few days if there is a `dark time' that occurs after the explosion \citep{2013ApJ...769...67P}. A shell that is sufficiently compact could interact and cool sufficiently that it is not detectable even within the first few hours after the explosion. There are generally two possible sources for such a shell --- a shell associated with accretion onto the progenitor white dwarf, or a shell of material ejected by a detonation on the surface of the white dwarf just prior to the explosion of the white dwarf itself (c.f. \citet{2010ApJ...715..767S} and references therein). The latter model will be nearly identical to the former so long as the material ejected in the surface detonation is moving sufficiently slowly that it is quickly overtaken by the outer edges of the ejecta from the final explosion.

In this work we consider compact ($R < $\Rsun) circumstellar shells as a source of the CaNIR HVF in SN~2011fe. SN~2011fe provides one of the earliest detections of a SN~Ia with associated spectra available to-date. We seek to evaluate the supernova--shell interaction models of \citet{2017MNRAS.467..778M} with a focus on the temporal evolution of the velocity, shape, and depth of the HVF. We fit spectra generated from the models to observational data of SN~2011fe from the earliest available epoch through to 9~d after B-band maximum. We seek to determine the plausibility of the supernova-shell interaction model, to identify the mass of this shell in SN~2011fe, and to understand how the opacity of the shell and ejecta evolve in the first weeks after the explosion.

In \S\ref{sec:methods} we describe the methods we use to generate synthetic spectra from the models and how we fit these models to observed spectra of SN~2011fe. In \S\ref{Sec:results} we describe the results of the fitting and investigate the observed trends in photosphere temperature and velocity. We discuss the methods used for determining velocity of the PVF and HVF in \S\ref{ssec:velev}. We discuss the impact of these results on some progenitor or explosion models in \S\ref{ssec:altmodel}. We summarize the results in \S\ref{sec:conclusion}. We have included a discussion of the challenges of fitting spectra in high order parameter space in Appendix \ref{sec:optfit}.

\section{Methods}\label{sec:methods}
\subsection{Observational data}
The observed spectra to which the models are fit were obtained through the Open Supernova Catalog \citep{2017ApJ...835...64G}. The sources of the observed spectra for each epoch are listed in Table \ref{parametertable}. We performed fits for data that are available in the first 4 days SN~2011fe was visible and include the CaNIR feature. This is the time period when HVF are strongest and PVF are weakest. After the first 4 days we use only data taken with the Hubble Space Telescope (HST) as these spectra have high signal-to-noise, there is no concern of telluric features and the data have a much larger wavelength range ($<3000$ -- $>10000\Ang$) than any ground-based spectra. The latter is especially important for flux scaling, described below, and determination of the photospheric temperature. We refer to the observed specific flux from any given spectral data set as $F^{\mathrm{obs}}_{\lambda}$. Because the relative strength of the HVF is dependent upon the time elapsed since the explosion, we hereafter refer to time as that relative to the explosion, taken to be $t_{\mathrm{exp}}=\mathrm{MJD}\;55796.696$ \citep{Nugent2011}. Before fitting, the spectra are shifted to the restframe wavelength, then dereddened according to \citet{1989ApJ...345..245C} with corrections to the visual range from \citet{1994ApJ...422..158O}. We adopt a value of E(B-V) = 0.0077 and a redshift of $z = 0.000804$ \citep{2014MNRAS.439.1959M}.

\begin{table*}
\caption{Data sources, fit ranges, and flux scaling ranges by date of observation \label{parametertable}}
\begin{center}
\begin{tabular}{lrrccc}
\hline
Observation & &Time since & Flux Scaling & Fit & SN~2011fe\\
Date & Phase$^{a}$ & explosion$^{b}$ & Range & Range & Data\\
$[\mathrm{MJD}]$ & $[\mathrm{days}]$ & $[\mathrm{days}]$ & $[\mathrm{\nuAng}]$ & $[\mathrm{\nuAng}]$ & Sources\\
\hline
55797.86 &	-17.04	& 1.16 & 3900 -- 9200 &	7680 -- 8900 & 1, 5\\
55798.3 &	-16.6	& 1.6 & 3400 -- 9300 &	7600 -- 8600 & 1, 5 \\
55799.26 &	-15.64	& 2.56 & 3290 -- 9700 &	8000 -- 8800 & 3, 5\\
55799.3 &	-15.6	& 2.6 & 3290 -- 9700 &	7790 -- 8700 & 2, 5\\
55801.17 &	-13.73	& 4.47 & 2700 -- 16800 &	7930 -- 8700 & 4, 5\\
55804.25 &	-10.65	& 7.55 & 2900 -- 16800 &	7970 -- 8750 & 4, 5 \\
55807.38 &	-7.52	& 10.68 & 1780 -- 16800 &	7930 -- 8700 & 4, 5 \\
55811.37 &	-3.53	& 14.67 & 1780 -- 16800 &	7910 -- 8750 & 4, 5 \\
55814.39 &	-0.51	& 17.69 & 1780 -- 16800 &	7930 -- 8700 & 4, 5 \\
55817.67 &	2.77		& 20.97 & 1700 -- 16500 &	7940 -- 8800 & 4, 5 \\
55823.62 &	8.72		& 26.92 & 1740 -- 10210 &	8000 -- 8800 & 4, 5 \\
\end{tabular}
\begin{flushleft}
${}^a$ Relative to B-band maximum on MJD 55814.90 \citep{2012MNRAS.426.2359M}.\\
${}^b$ Based on explosion at MJD 55796.696 \citep{Nugent2011}.\\
1: \citet{Nugent2011}\\
2: \citet{2012ApJ...752L..26P}\\
3: \citet{2013AnA...554A..27P}\\
4: \citet{2014MNRAS.439.1959M}\\
5: \citet{2017ApJ...835...64G}\\
\end{flushleft}
\end{center}\end{table*}

\subsection{Supernova-shell interaction models}
In \citet{2017MNRAS.467..778M} we reported the results of hydrodynamic simulations of the interaction bewteen a SN~Ia and compact, circumstellar shells. The shells have a range of initial radii between 0.04\Rsun and 1\Rsunp, masses between 0.001\Msun and 0.02\Msun, and different initial density profiles, equations of state, and underlying explosion models. We found that of these parameters, the mass of the shell has the largest effect on the velocity and equivalent width of the CaNIR feature while preserving shapes of the feature that are commensurate with those that are observed. In this work, we use supernova-shell interaction model numbers 17, 41, 45, 49, 57, and 61, representing a supernova without interaction (model 17), and those interacting with shells with mass 0.003\Msun (model 61), 0.005\Msun (model 57), 0.008\Msun (model 41), 0.010\Msun (model 49) and 0.012\Msun (model 45). Each model has a shell with and initial outer radius of 0.3\Rsunp, uses \citet{Gamezo2005} delayed detonation model \texttt{c} for the explosion and ejecta with the gamma-law equation of state (excepting model 17, which uses a Helmholtz equation of state), and has a sawtooth density profile with the highest density at the edge closest to the explosion. Hereafter we refer to supernova-shell interaction models by the mass of the shell rather than model number for clarity.

The models provide the spatially resolved density and velocity of the supernova and shell material, and, for the ejecta, the abundance of interesting elements (e.g. C, O, Si, etc.). The hydrodynamic data is used to produce a normalized, dimensionless density profile ($\mathcal{G}^c_{v,i}$) for each ion $i$ and component $c$ (i.e. the ejecta, E,  and the shell, S) as a function of velocity ($v$). The profile is described over 2048 velocity bins spanning the range of velocity for all material for each model, and normalized by the spatial maximum of the total gas density (for the shell) or the density of a given element (for the ejecta).

\subsection{Synthetic Spectra}\label{sec:syn_spectra}

In generating synthetic spectra, we are interested in only one absorbtion feature (CaNIR), which we presume to be unblended with any other features. In addition, we do not assume an \textit{a priori} composition of the material within the shell. As such, we use a modified version\footnote{The modified version of \textsoft{syn++} is publicly available on github.com in repository \texttt{astrobit/es}.} of \textsoft{syn++} \citep{2011PASP..123..237T} that allows for selection of individual ions and an arbitrary profile of optical depth versus velocity.

The optical depth at velocity $v$ for ion $i$ is given by
\begin{equation}\tau_{v,i}=\sum_{c}\left(S^c_i\mathcal{G}^c_{v,i}\right),\end{equation}
where $S^c_i$ is a scalar factor for each ion and component that may contribute to absorption within the supernova. The scalar factors are a proxy for elemental abundance, ion- and excitation-states, and line specific components of the line optical depths. When generating a spectrum with \textsoft{syn++}, the temperature and velocity of the photosphere must also be specified. These set the shape of the blackbody continuum and the minimum velocity in the profile ($\mathcal{G}_{v,i}$) that has an effect on the spectrum. 

The spectra generated by \textsoft{syn++} have a relative flux value between 0 - 1, whereas the observational data may be reported in $\mathrm{counts}\:\mathrm{sec}^{-1}$ or $\mathrm{erg}\:\mathrm{sec}^{-1}\Ang^{-1}\:\mathrm{cm}^{-2}$. We therefore scale the relative flux such that the total flux over some range of wavelengths matches the observed flux over the same range for a given observation. The resulting scaled synthetic flux is then 
\begin{equation}\label{eq:flux_scaling}F^{\mathrm{syn}}_{\lambda}=\left(\frac{\sum_{\lambda}F^{\mathrm{obs}}_\lambda}{\sum_{\lambda}f^{\mathrm{syn}}_\lambda}\right)f^{\mathrm{syn}}_{\lambda},\end{equation}
where $f^{\mathrm{syn}}_\lambda$ is the relative flux generated by \textsoft{syn++}, and the sums are over the flux scaling ranges listed in Table \ref{parametertable}.

\subsection{\label{ssec:fitting}Fitting}
We define the parameter space for fitting with four parameters: temperature and velocity of the photosphere ($T_{\mathrm{PS}}$ and $v_{\mathrm{PS}}$) and the log of the scalar factors $S^\mathrm{E}_{\ion{Ca}{II}}$ and $S^\mathrm{S}_{\ion{Ca}{II}}$ for the ejecta and shell components, respectively. We collectively describe the parameters as vector $\bm{x}=[T_{\mathrm{PS}},v_{\mathrm{PS}},\log S^\mathrm{E}_{\ion{Ca}{II}},\log S^\mathrm{S}_{\ion{Ca}{II}}]$. A starting point is chosen that has a reasonably good fit by eye, then a grid search and refinement is performed. At each step, the grid consists of a hypercube of $3^4$ vertices in parameter space, i.e. 3 values for each parameter. The initial grid has a total range of $\bm{\Delta x} =$ $[10000\Kelvin,5000\kms,0.4,0.4]$. After generating spectra at each vertex in the grid, we identify the vertex that minimizes the variance,
\begin{equation}J=\frac{1}{2 N} \sum_{\lambda=\lambda^\mathrm{Fit}_\mathrm{min}}^{\lambda^\mathrm{Fit}_\mathrm{max}}\left(F^{\mathrm{obs}}_{\lambda}-F^{\mathrm{syn}}_{\lambda}(\bm{x})\right)^2,\end{equation} 
where N is the number of data points in the selected fitting range. The range of wavelengths over which each spectrum is fit are listed in Table \ref{parametertable}. The fitting ranges include most of the P~Cygni emission component on the red side of the feature and the identifiable blue side of the feature (including HVF), excluding regions potentially affected by P~Cygni emission of the $7500\Ang$ feature blue-ward of the CaNIR feature. The vertex with the best fit is then chosen as the next starting point and the size of the grid is halved. This process is iterated to a final grid size of $\bm{\Delta x}=[19\Kelvin,10\kms,0.0008,0.0008]$. The use of variance to find the best fitting set of parameters and model does not fully capture the differences in shape of the feature; further, the use of an adaptively-refined grid search can result in an identification of a local, rather that global minima, although for this parameter space we find that the topography seems relatively smooth. We discuss these issues further in Appendix \ref{sec:optfit}.

\section{Results and discussion}\label{Sec:results}
The variances of the fits by model and time after the explosion are listed in Table \ref{tab:variance}. No single model clearly has the best quality of fit over all epochs, but the model with no shell performs poorly prior to 8~d, when the feature is dominated by HVF. The fits between models with a shell are generally of similar quality, though the model with a shell of mass 0.005\Msun tends to be the best fit at more epochs than any other model.

\begin{table*}\begin{center}
\caption{Variance of fit by time since explosion and mass of the shell\label{tab:variance}}
\begin{tabular}{rcrrrrrr}
Time Since & & & & & & & \\
Explosion $[\mathrm{d}]$ & Units&No Shell & 0.003\Msun & 0.005\Msun & 0.008\Msun & 0.010\Msun & 0.012\Msun\\
\hline
1.16 & $10^{-33}\:\mathrm{erg}^2\:\mathrm{cm}^{-4}\:\mathrm{sec}^{-2}$ & 19.42 & 6.92 & 7.00 & 7.04 & 7.10 & 7.22 \\
1.60 & $10^{-3}\:\mathrm{counts}^2\:\mathrm{sec}^{-2}$ & 136.62 & 4.03 & 3.58 & 7.12 & 5.55 & 5.42 \\
2.56 & $10^{-32}\:\mathrm{erg}^2\:\mathrm{cm}^{-4}\:\mathrm{sec}^{-2}$ & 12.85 & 10.85 & 11.79 & 10.63 & 10.44 & 9.87 \\
2.60 & $10^{-32}\:\mathrm{erg}^2\:\mathrm{cm}^{-4}\:\mathrm{sec}^{-2}$ & 102.90 & 4.33 & 18.46 & 3.56 & 26.24 & 27.27 \\
4.47 & $10^{-31}\:\mathrm{erg}^2\:\mathrm{cm}^{-4}\:\mathrm{sec}^{-2}$ & 4.79 & 1.04 & 1.02 & 1.92 & 1.23 & ... \\
7.55 & $10^{-31}\:\mathrm{erg}^2\:\mathrm{cm}^{-4}\:\mathrm{sec}^{-2}$ & 55.92 & 13.77 & 3.28 & 4.16 & 8.21 & 11.35 \\
10.68 & $10^{-30}\:\mathrm{erg}^2\:\mathrm{cm}^{-4}\:\mathrm{sec}^{-2}$ & 8.48 & 3.31 & 3.68 & 5.02 & 5.33 & 5.70 \\
14.67 & $10^{-30}\:\mathrm{erg}^2\:\mathrm{cm}^{-4}\:\mathrm{sec}^{-2}$ & 12.07 & 8.40 & 6.43 & 8.32 & 10.10 & 11.65 \\
17.69 & $10^{-30}\:\mathrm{erg}^2\:\mathrm{cm}^{-4}\:\mathrm{sec}^{-2}$ & 17.04 & 14.36 & 7.32 & 17.93 & 10.41 & 17.49 \\
20.97 & $10^{-30}\:\mathrm{erg}^2\:\mathrm{cm}^{-4}\:\mathrm{sec}^{-2}$ & 7.14 & 7.36 & 8.62 & 7.86 & 8.59 & 13.19 \\
26.92 & $10^{-30}\:\mathrm{erg}^2\:\mathrm{cm}^{-4}\:\mathrm{sec}^{-2}$ & 11.20 & 8.97 & 9.50 & 9.82 & 11.19 & 13.61 \\
\end{tabular}\end{center}\end{table*}

\begin{figure}\centering
\includegraphics[width=0.37\textwidth,angle=-90]{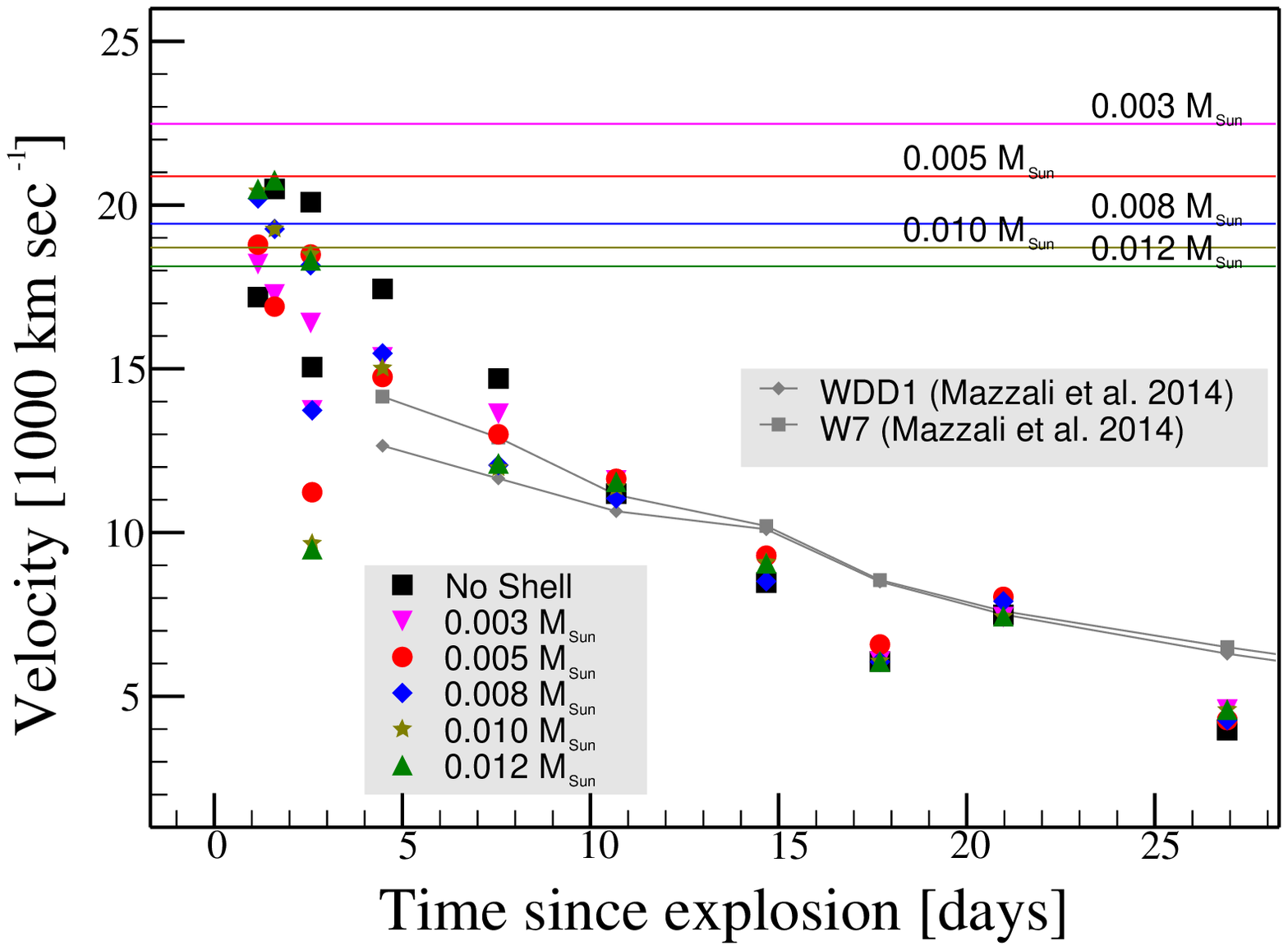}
\caption{\label{fig:ps_vel_evol} The evolution of the velocity of the photosphere for each model as well as the photosphere velocities as estimated in \citet{2014MNRAS.439.1959M} based on the W7 and WDD1 \citep{1984ApJ...286..644N,1999ApJS..125..439I} supernova models. For models with a shell, the velocity of the contact discontinuity is shown as colored lines, labelled just above each line. The photosphere is at lower velocity than the contact discontinuity at all epochs except for those of the highest mass shells. Uncertainties (not shown) are estimated to be less than $100\kms$ for all points.}
\end{figure}

Figure \ref{fig:ps_vel_evol} shows the velocity of the photosphere for each model as well as the velocity of the photosphere reported in \citet{2014MNRAS.439.1959M}, determined using the W7 and WDD1 \citep{1984ApJ...286..644N,1999ApJS..125..439I} models. The photosphere velocity decreases monotonically for each model from about 4 days after explosion to about maximum light. There is a larger scatter among models in the velocity of the photosphere at early epochs (4~d and earlier). During this period, HVF are relatively strong and the fit is more dependent upon the material in the shell. After 8~d the CaNIR feature is dominated by absorption within the ejecta and thus the velocity of the photosphere is independent of the details of the shell. The photosphere lies within the shell in the first several days for the models with a shell of mass 0.008\Msun or larger; in the shells of lower mass (0.003\Msun and 0.005\Msunp) the photosphere is in the ejecta at all times. This suggests that for shells of higher mass there could be broadband photometric signatures of the shell in the first several days after the explosion. Hydrodynamic simulations that include cooling effects are required to understand the temperature, and thus the luminosity, of the shell at these epochs. A broadband photometric signature will only occur for a period of minutes for shells of lower mass, satisfying the limits of \citet{2012ApJ...744L..17B}, though there may be line emission for a longer period.

\begin{figure}\centering
\includegraphics[width=0.37\textwidth,angle=-90]{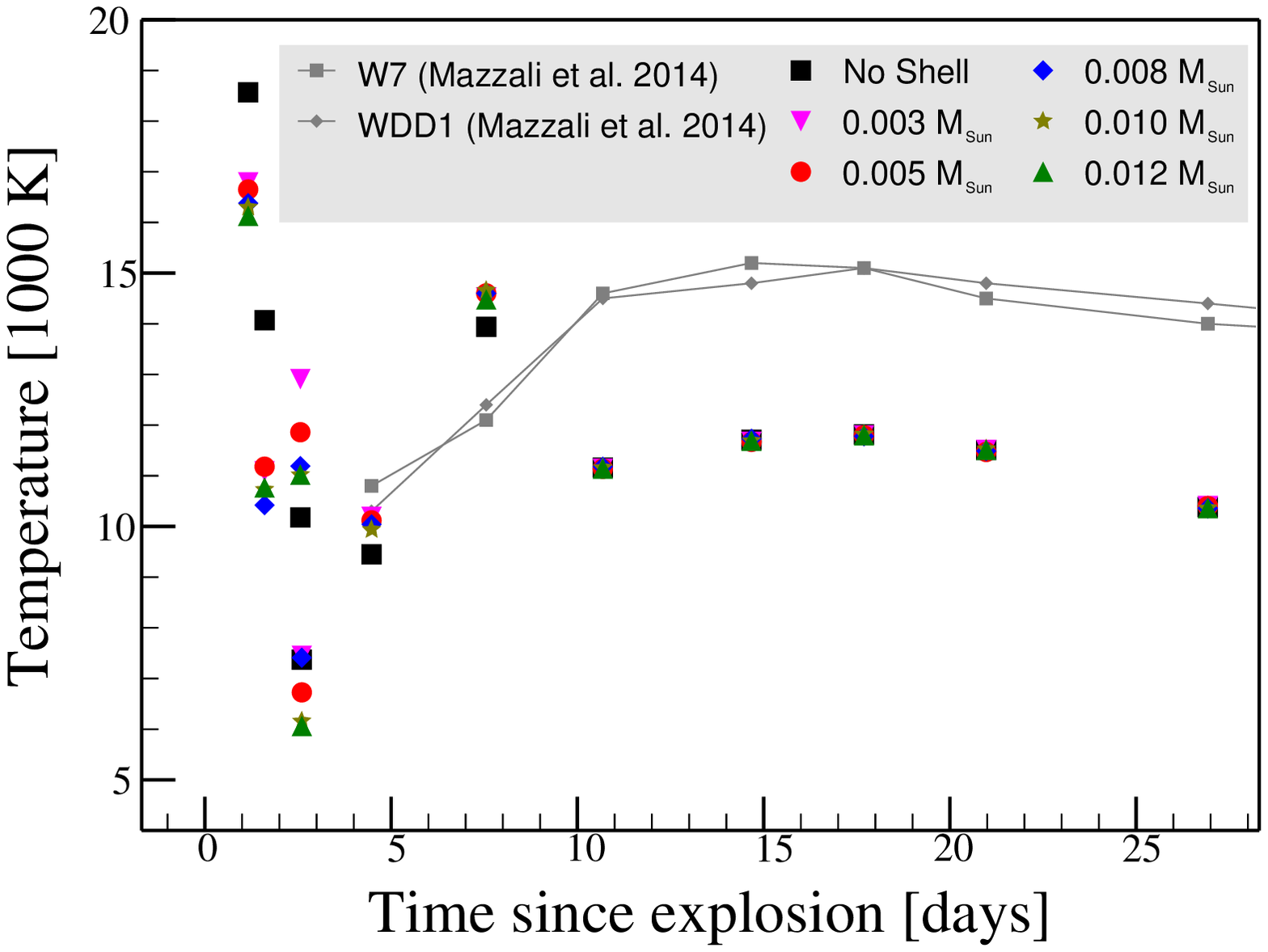}
\caption{\label{fig:ps_temp_evol} The evolution of the temperature of the photosphere, along with the photosphere temperatures estimated in \citet{2014MNRAS.439.1959M} based on the W7 and WDD1 \citep{1984ApJ...286..644N,1999ApJS..125..439I} supernova models. The photosphere temperatures are consistent with expected supernova photosphere temperatures, but tend to be about 4000~K cooler than those estimated in \citet{2014MNRAS.439.1959M}. The scatter in the first four days is due to the limited range available for flux scaling (see \S\ref{sec:syn_spectra}). Uncertainties (not shown) are estimated to be less than $100\Kelvin$ for all points.}
\end{figure}

The temperature of the photosphere, shown in Figure \ref{fig:ps_temp_evol}, determines the specific flux and slope of the continuum in the vicinity of the CaNIR feature. These temperatures are about 4000~K lower than those found by \citet{2014MNRAS.439.1959M}, that utilise a more complete set of spectral features, though the overall trend is similar.
The ground-based data at 2.5~d and earlier have a smaller wavelength range over which the flux scaling can be done compared to the spectra obtained with HST. The variation in temperature at these epochs is because the flux scaling does not capture the effect of features in the near-ultraviolet or near-infrared. 

\begin{figure}\centering
\includegraphics[width=0.37\textwidth,angle=-90]{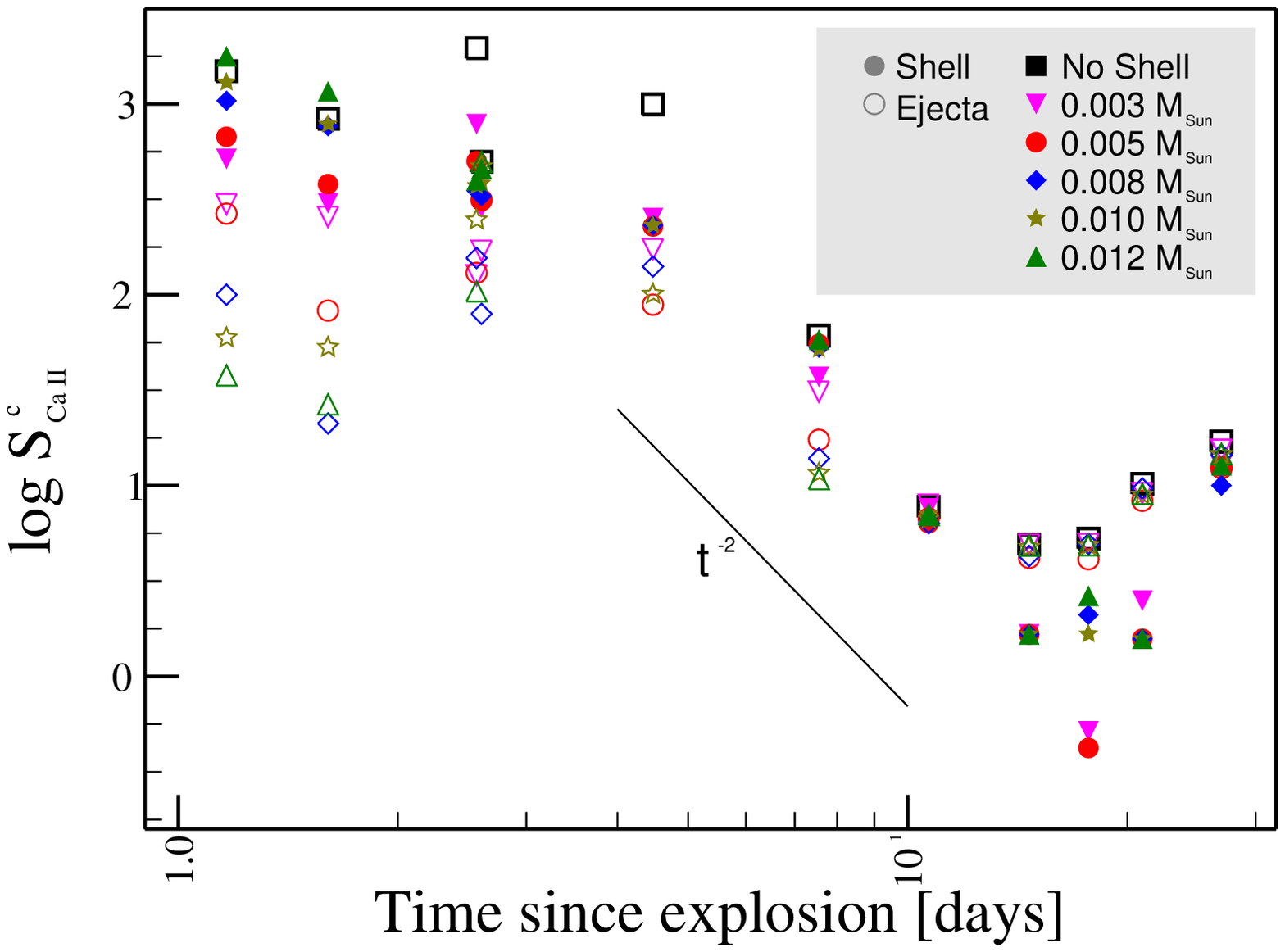}
\caption{\label{fig:scalar_evol} The evolution of the scalar factors $S^c_{\ion{Ca}{II}}$ for each model. The $t^{-2}$ trend in scalars expected for material in free expansion with constant ionization and excitation states is shown as a grey line. No individual models follow this trend for more than a few days. See \S\ref{Sec:results} for further details and discussion. Uncertainties (not shown) are estimated to be less than 0.01~dex for all points.}
\end{figure}

As illustrated in Figure \ref{fig:scalar_evol}, the scalar factors ($\mathrm{log}S^c_{\ion{Ca}{II}}$) arising from the fits tend to decrease until about the time of Bmax, then begin to increase. The shell component is weak after about 7~d, so the overall fit is weakly dependent on the scalar factor for the shell. The late-time increase in the scalar factor for the ejecta, however, may be an indicator that there is more calcium at lower velocities than there is in the explosion model. Alternatively, the relative strength of absorption may be changing because of an increasing population in the ejecta of the \ion{Ca}{II} in the lower state of the CaNIR transition.

For a gas in free expansion that has populations of all ionization and excitation states that vary slowly with respect to the expansion timescale, the scalar factors are expected to scale with density and decrease as $t^{-2}$. A line is given in Figure \ref{fig:scalar_evol} to demonstrate the expected trend. None of these models follow a $t^{-2}$ decline at any epoch, suggesting that the total populations of \ion{Ca}{II} in the lower state of the CaNIR transition are changing or the opacity profile that we have used does not match the profile within the supernova. The latter may be due to a non-uniform distribution of \ion{Ca}{II} in the shell (i.e. non-uniform abundance of calcium or variation of ionization or excitation state of calcium through the shell) and to differences between the structure of the ejecta in the explosion model and that of the supernova.

\begin{figure*}\centering\subfigure[No Shell]{\includegraphics[width=0.33\textwidth,angle=-90]{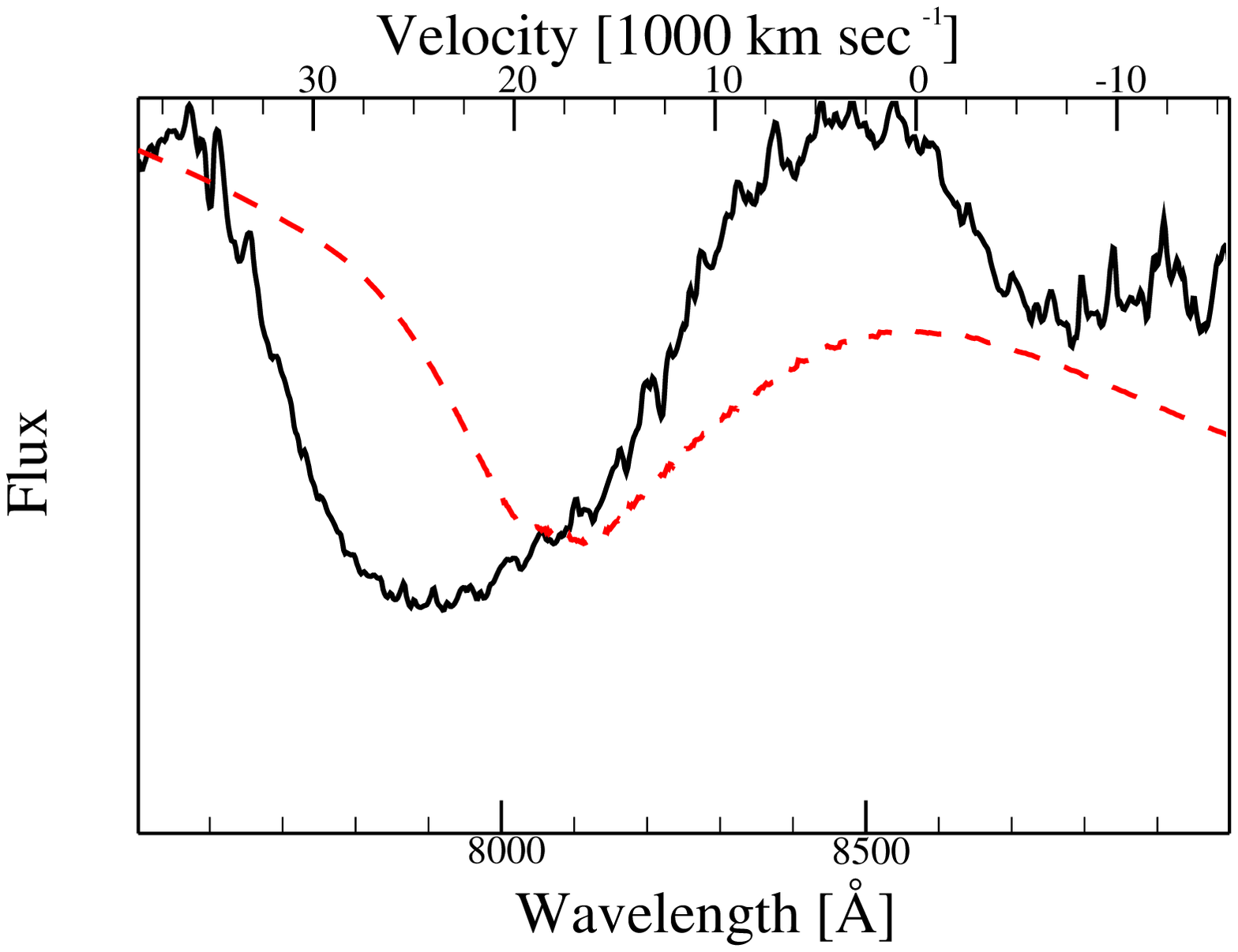}}\subfigure[0.003\Msun]{\includegraphics[width=0.33\textwidth,angle=-90]{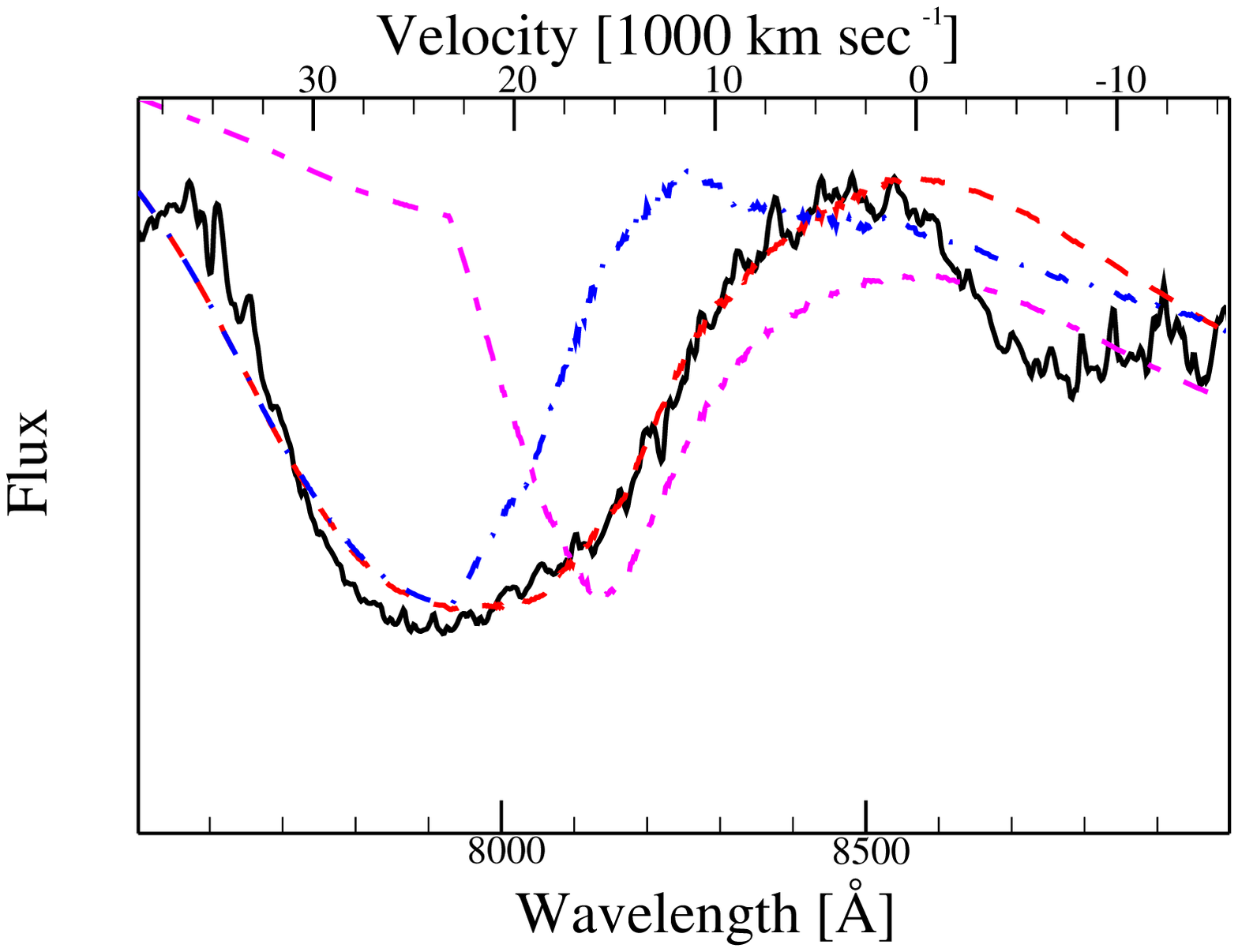}}
\subfigure[0.005\Msun]{\label{Fig2legend}\includegraphics[width=0.33\textwidth,angle=-90]{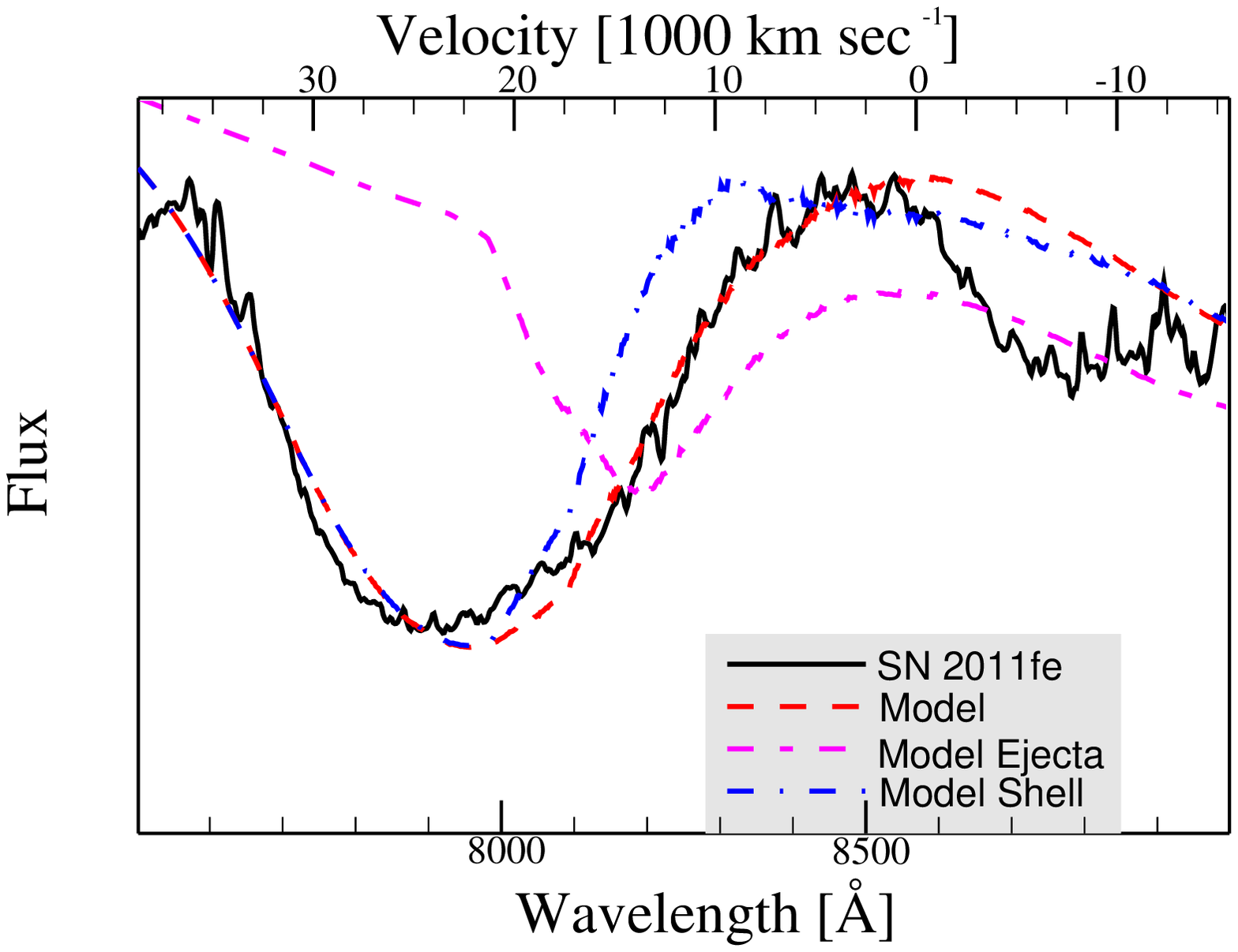}}\subfigure[0.008\Msun]{\includegraphics[width=0.33\textwidth,angle=-90]{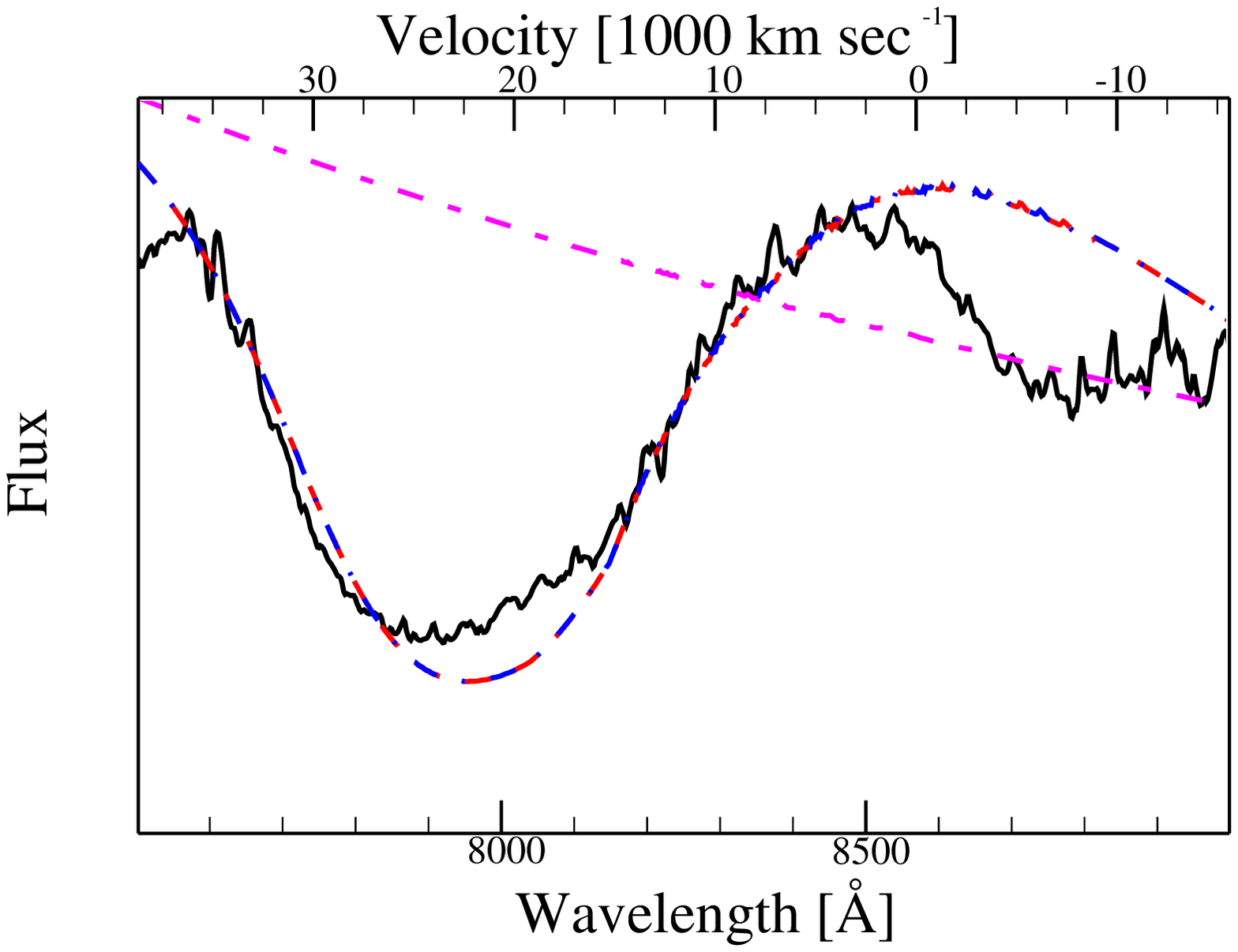}}
\subfigure[0.010\Msun]{\includegraphics[width=0.33\textwidth,angle=-90]{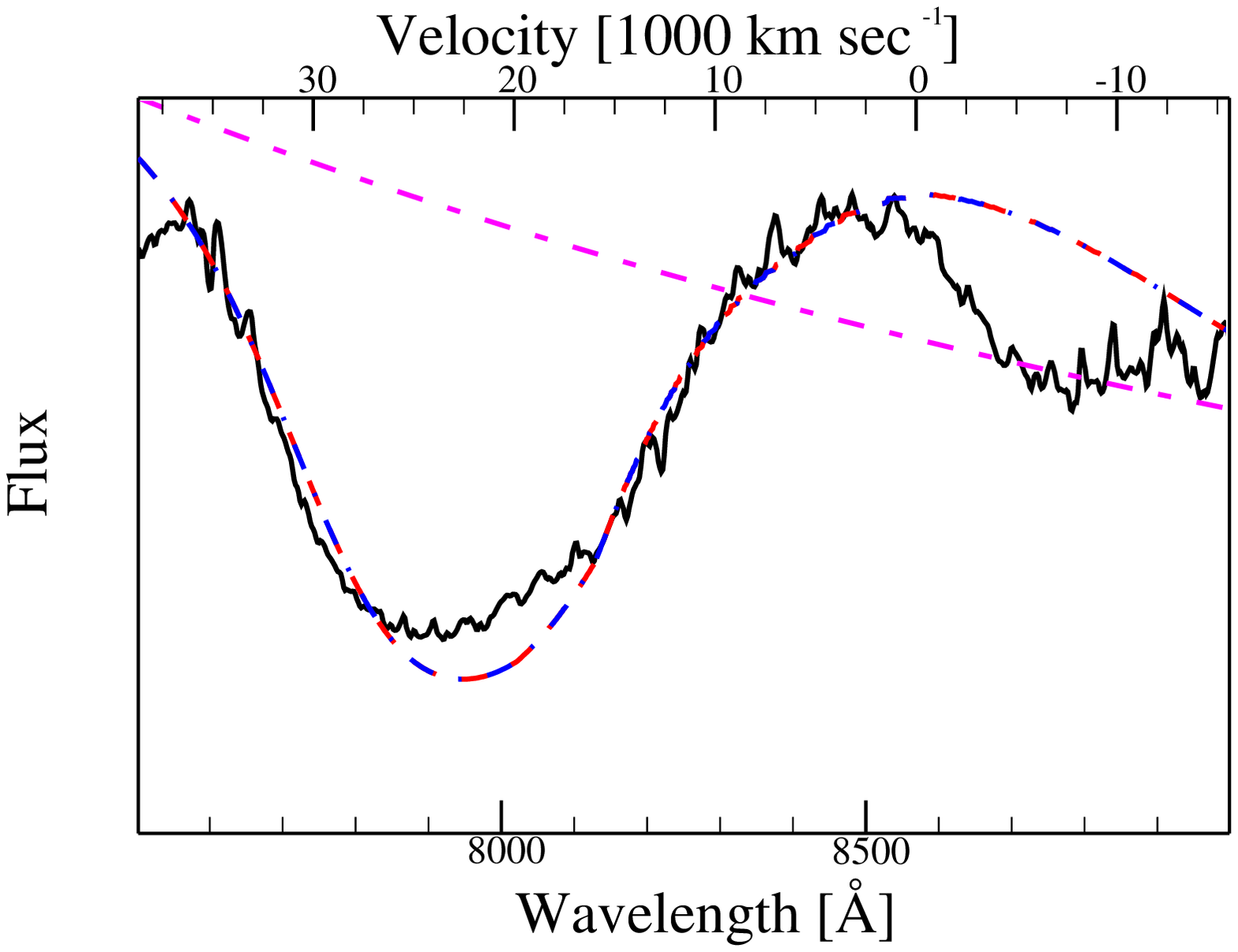}}\subfigure[0.012\Msun]{\includegraphics[width=0.33\textwidth,angle=-90]{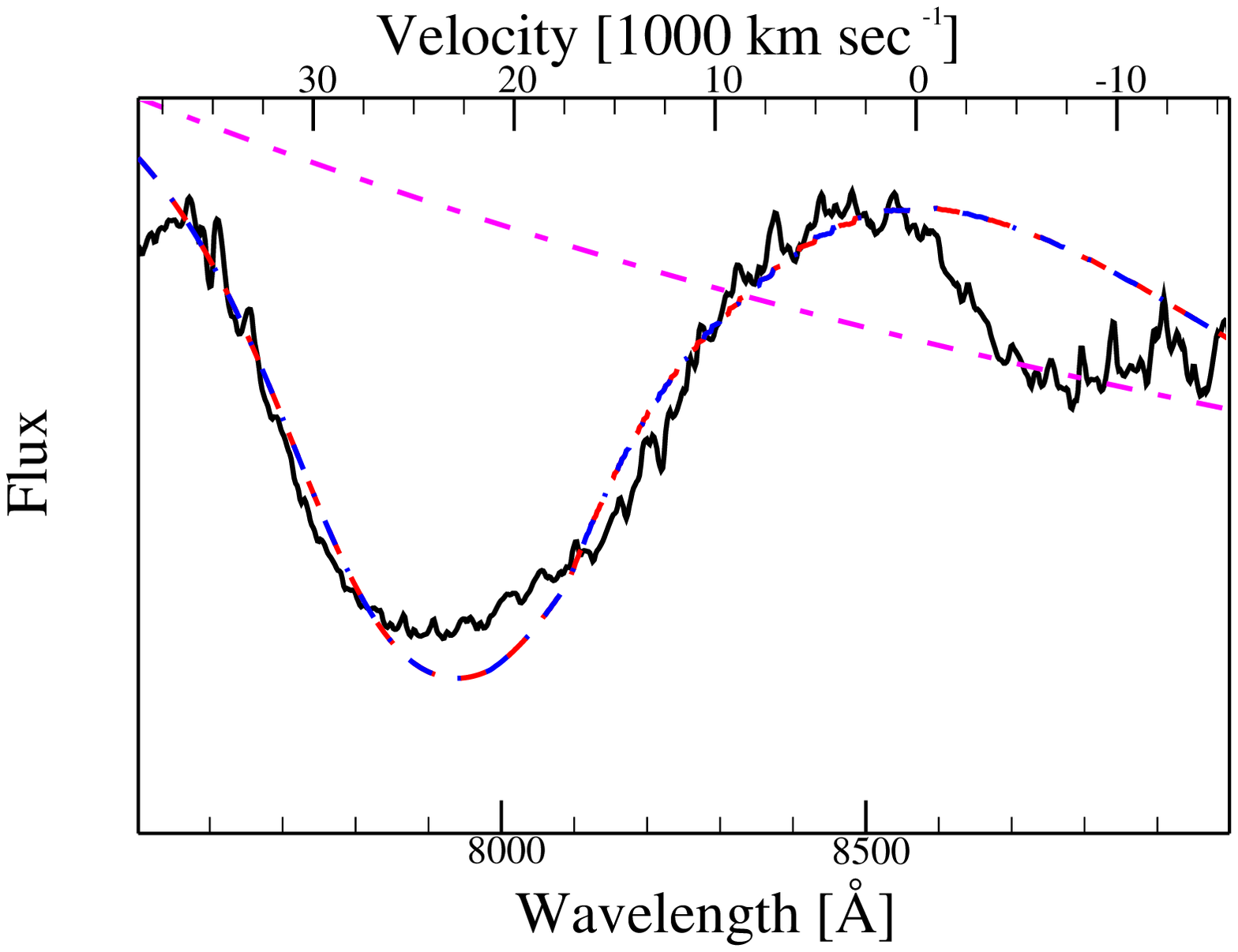}}
\caption{\label{Fig:allmodels55798} Results of fitting the CaNIR feature for all models to SN~2011fe at 1.6~d. SN~2011fe spectra are shown in black and the generated spectrum from the model is shown in dashed red. SN~2011fe data sources are listed in Table \ref{parametertable}. A legend is shown in Figure \ref{Fig2legend}. The shell with mass 0.005\Msun has the best quality of fit (see Table \ref{tab:variance}) and the model without a shell is an extremely poor fit at this epoch. The velocity scale is based on the central component of the CaNIR triplet. The magenta long-dash--short-dash line shows the flux when the material in the shell is excluded, and the blue long-dash--dotted line shows the flux when the material in the ejecta is excluded. These demonstrate the relative contributions of the ejecta and shell components in each model. The ejecta has no effect on the shape of the feature at this epoch for models with a shell of mass 0.008\Msun and larger.}\end{figure*}

Observed and synthetic spectra for all models at 1.6~d are shown in Figure \ref{Fig:allmodels55798}. The fit to the observed feature for the spectrum generated with the model with no shell is very poor at this epoch, while the quality of fit for all models with a shell tend to be similarly good. Each panel in Figure \ref{Fig:allmodels55798} shows a decomposed synthetic spectrum revealing the relative contributions of the shell and ejecta components. The models with a shell of mass 0.008\Msun or larger show only a continuum for the ejecta component indicating that only the shell has an effect on the CaNIR feature at this epoch, whereas the models with a shells of mass 0.003\Msun or 0.005\Msun show contributions from both the shell and ejecta components. This reflects that the photosphere lies within the shell at this epoch for the shells with highest mass, and that the photosphere is at lower velocity than the contact discontinuity for the lower mass models.

\begin{figure*}\centering\subfigure[1.16~d]{\includegraphics[width=0.25\textwidth,angle=-90]{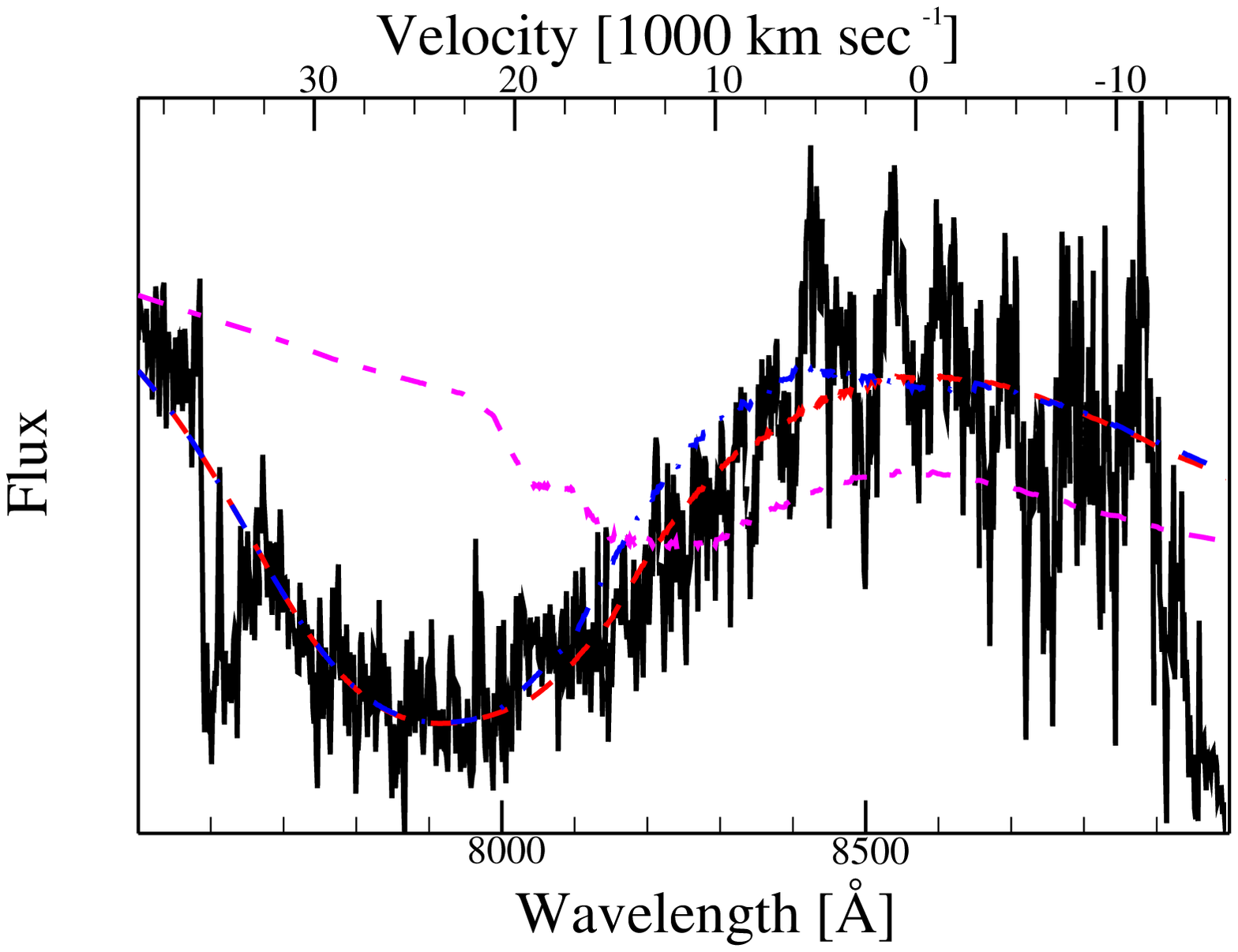}}\subfigure[1.6~d]{\includegraphics[width=0.25\textwidth,angle=-90]{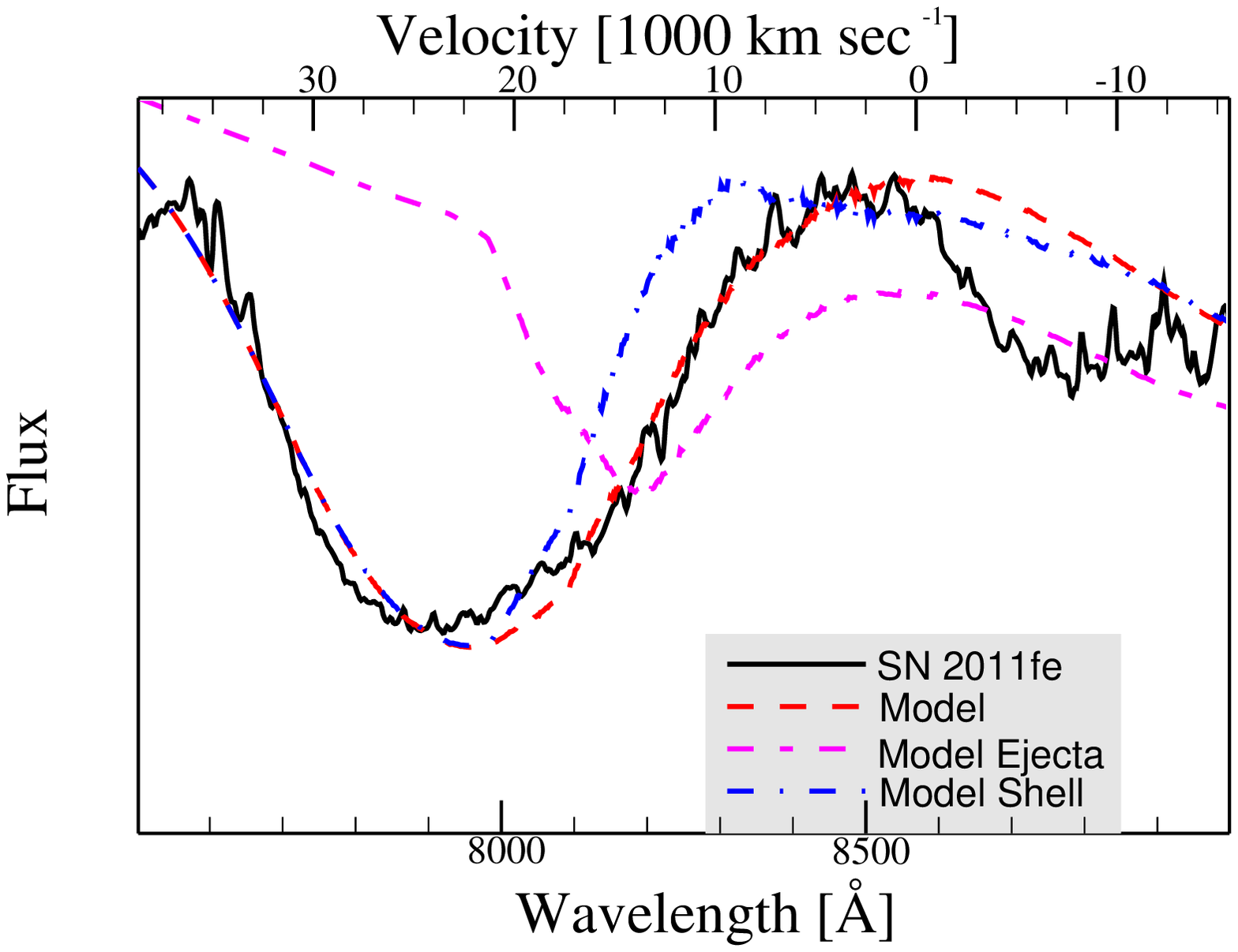}}\subfigure[2.56~d]{\includegraphics[width=0.25\textwidth,angle=-90]{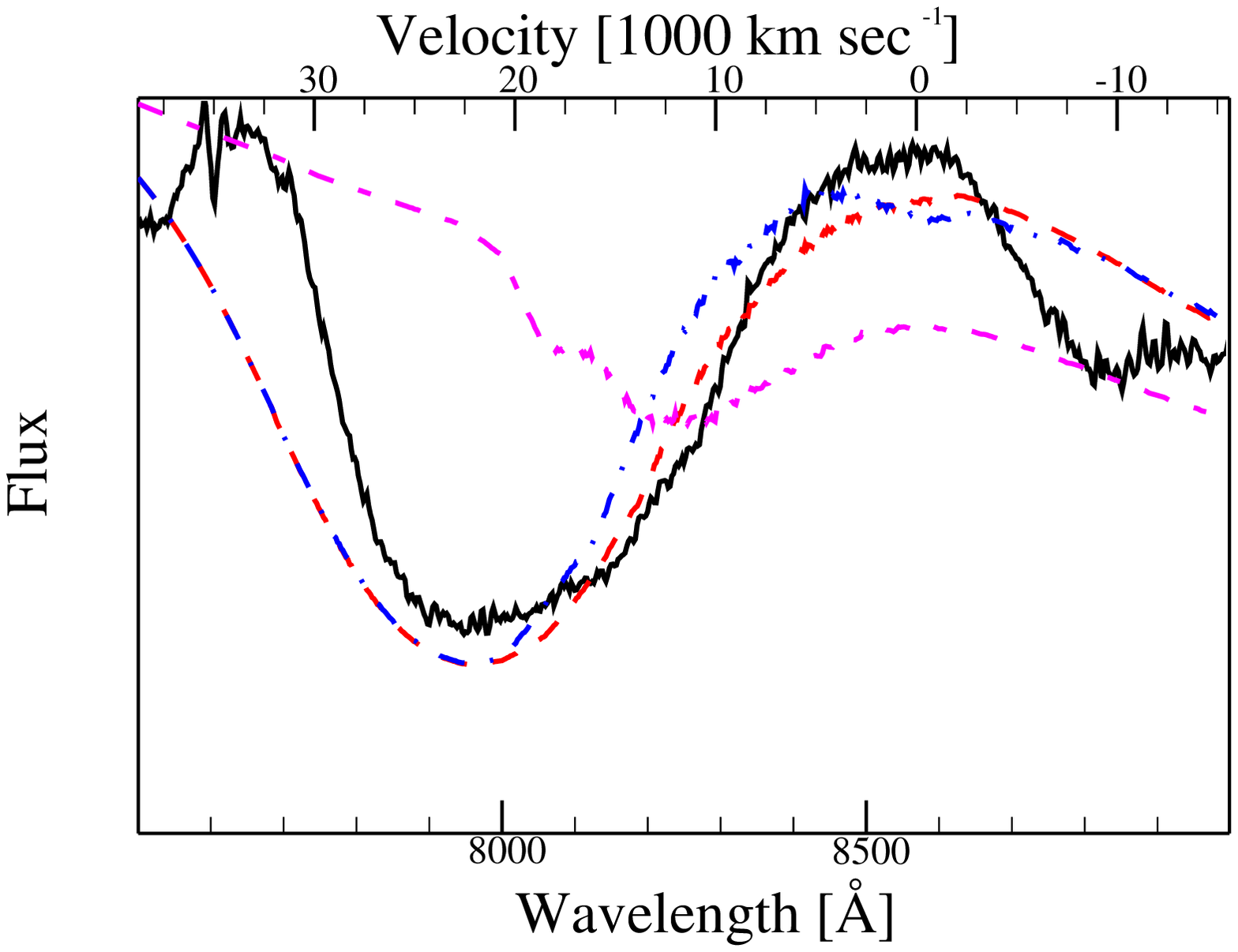}}
\subfigure[2.6~d]{\includegraphics[width=0.25\textwidth,angle=-90]{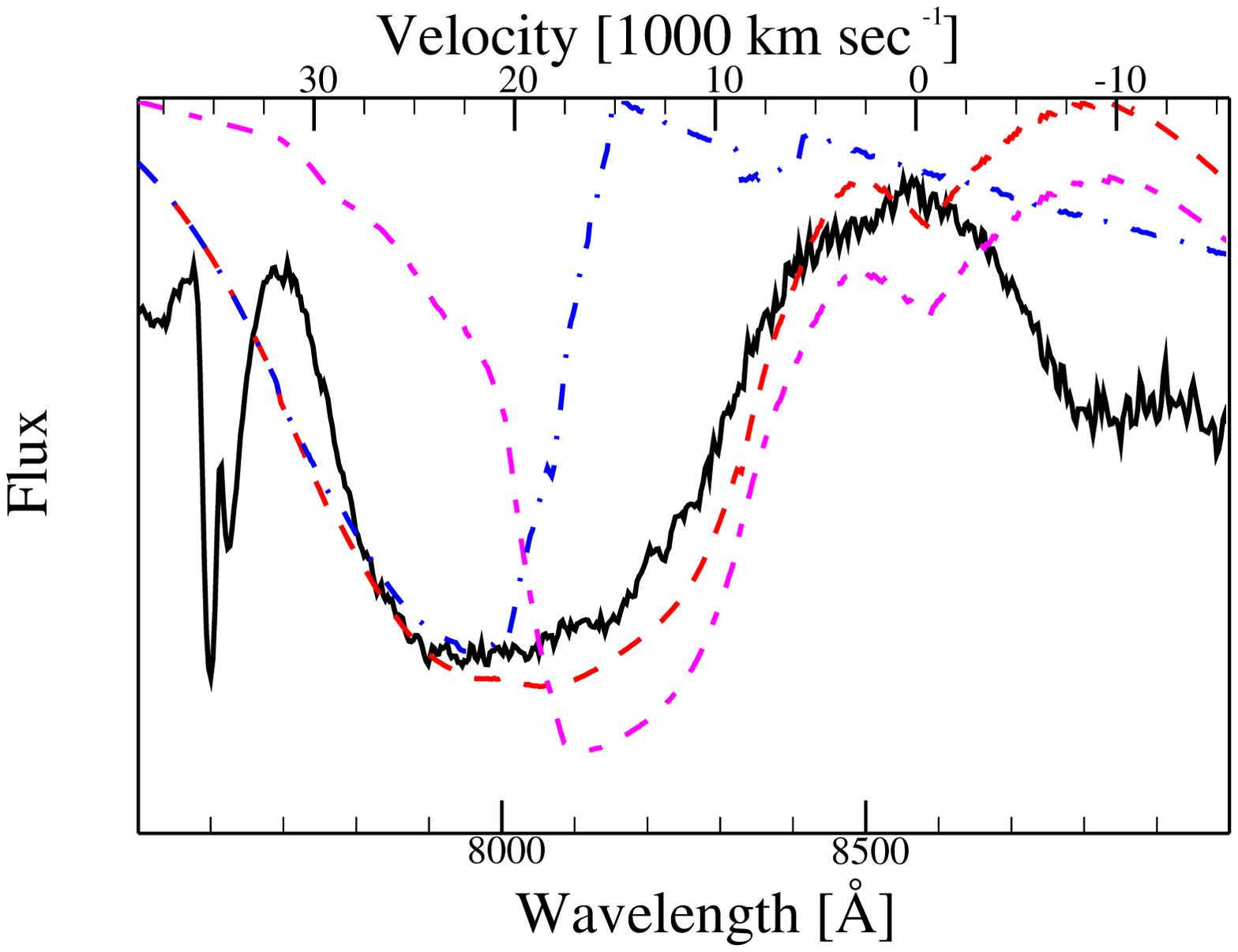}}\subfigure[4.4~d]{\includegraphics[width=0.25\textwidth,angle=-90]{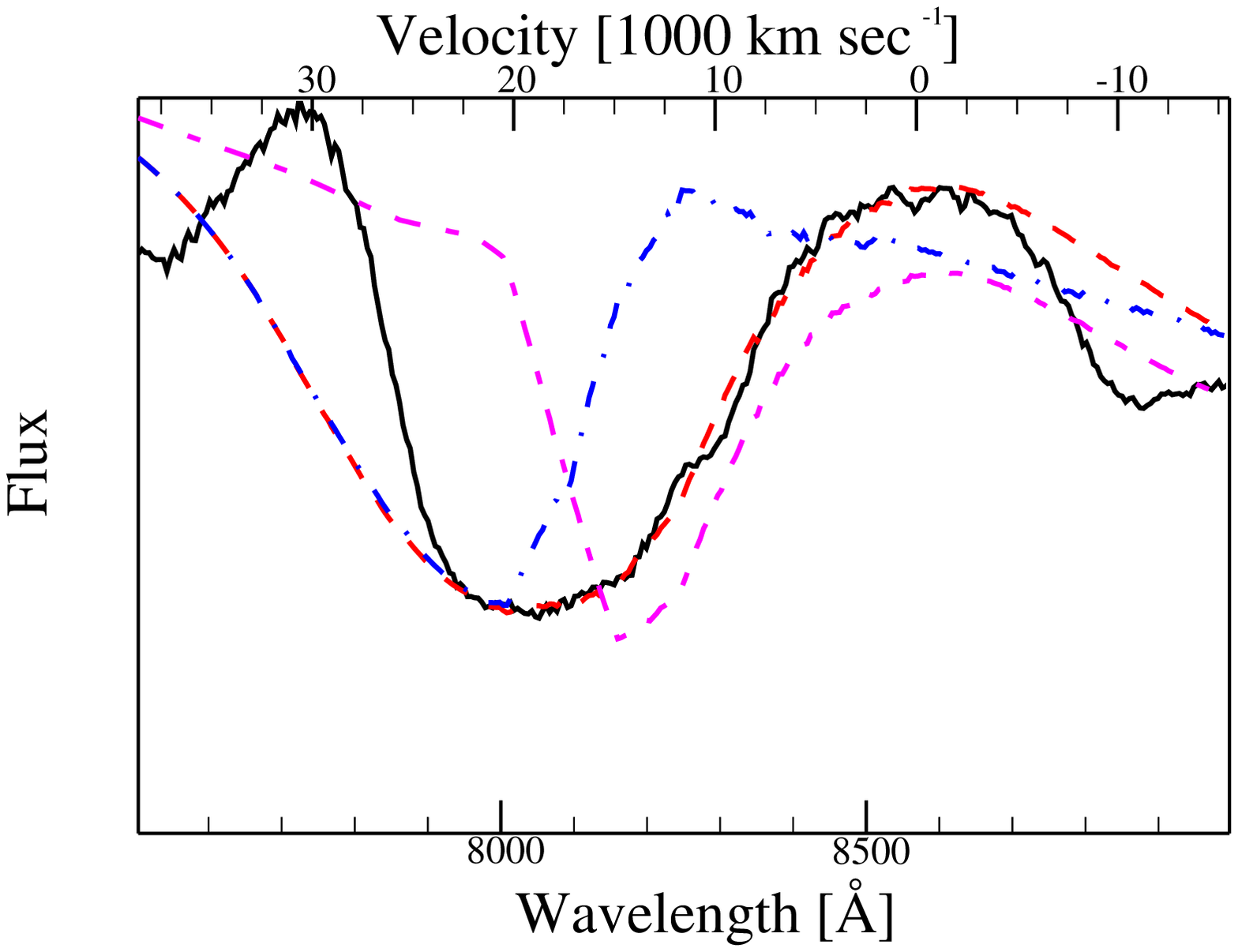}}\subfigure[7.5~d]{\includegraphics[width=0.25\textwidth,angle=-90]{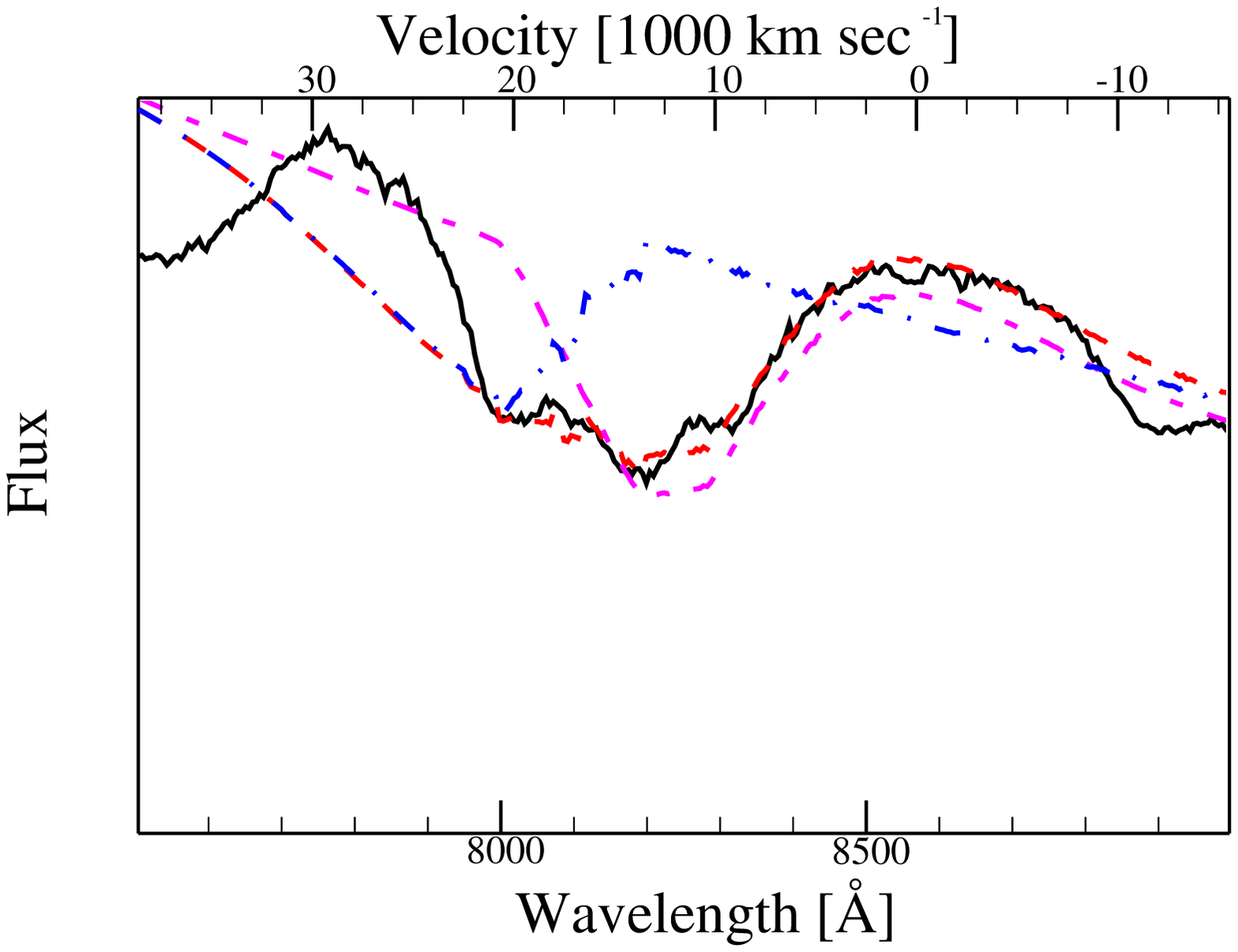}}
\subfigure[10.7~d]{\includegraphics[width=0.25\textwidth,angle=-90]{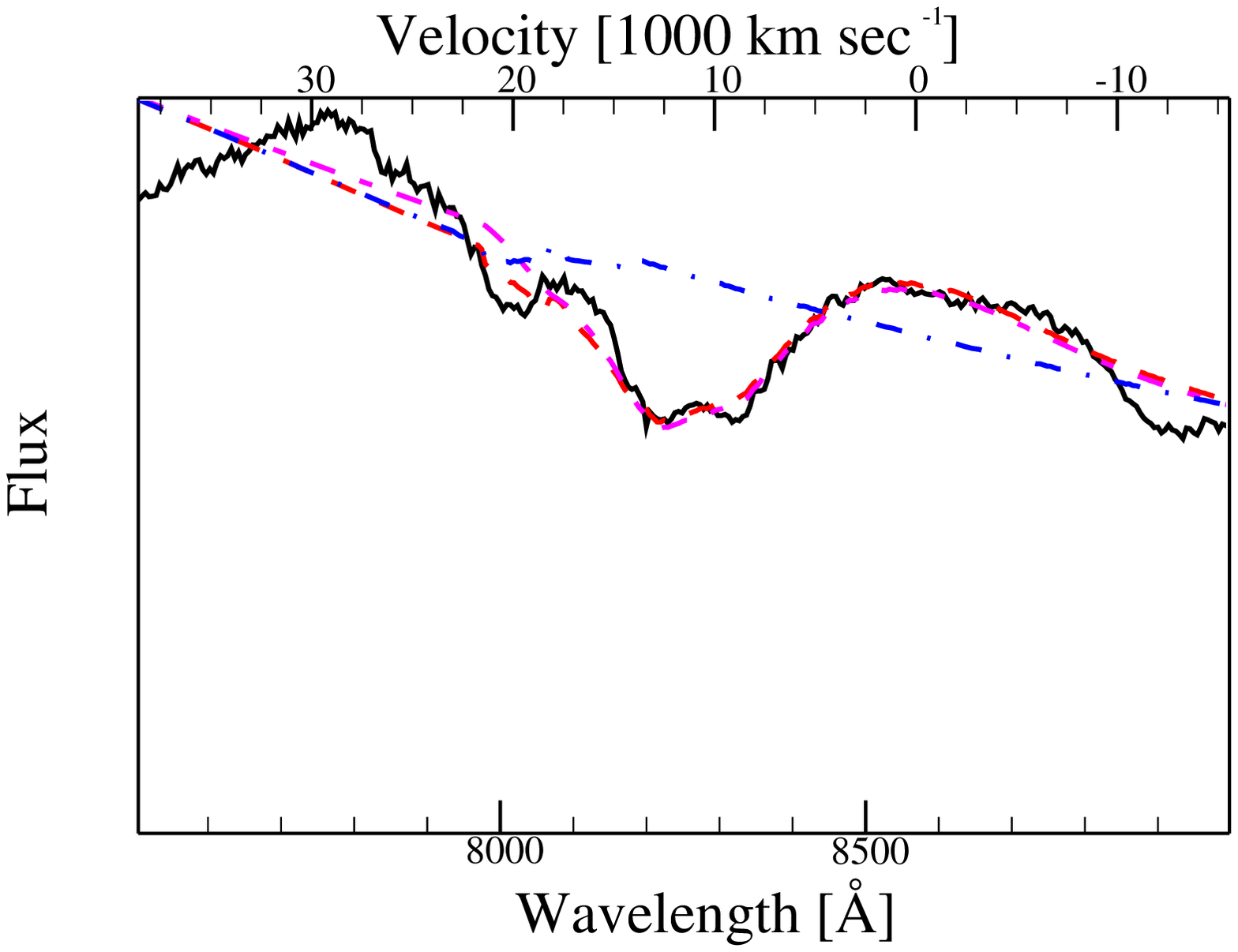}}\subfigure[14.7~d]{\includegraphics[width=0.25\textwidth,angle=-90]{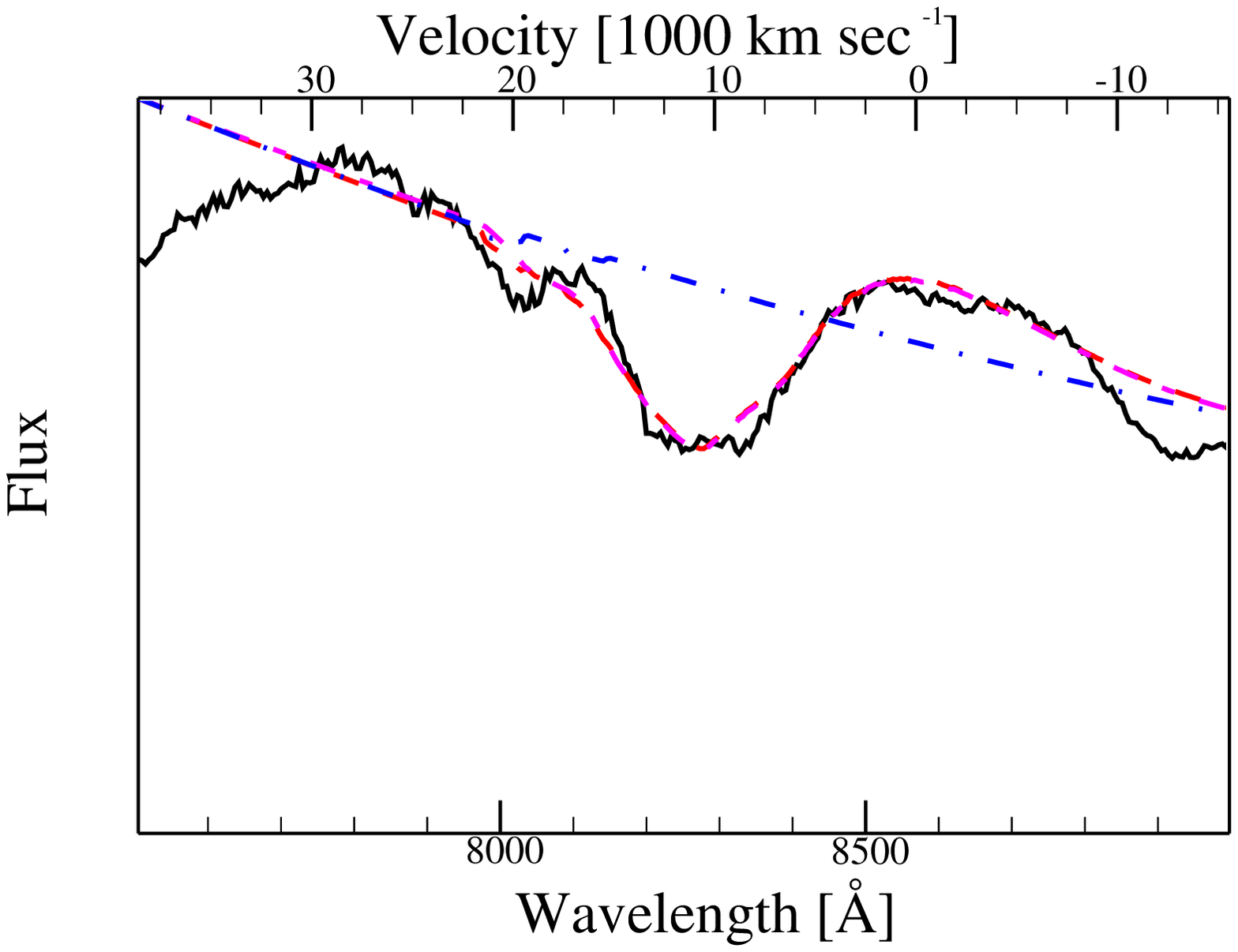}}\subfigure[17.7~d]{\includegraphics[width=0.25\textwidth,angle=-90]{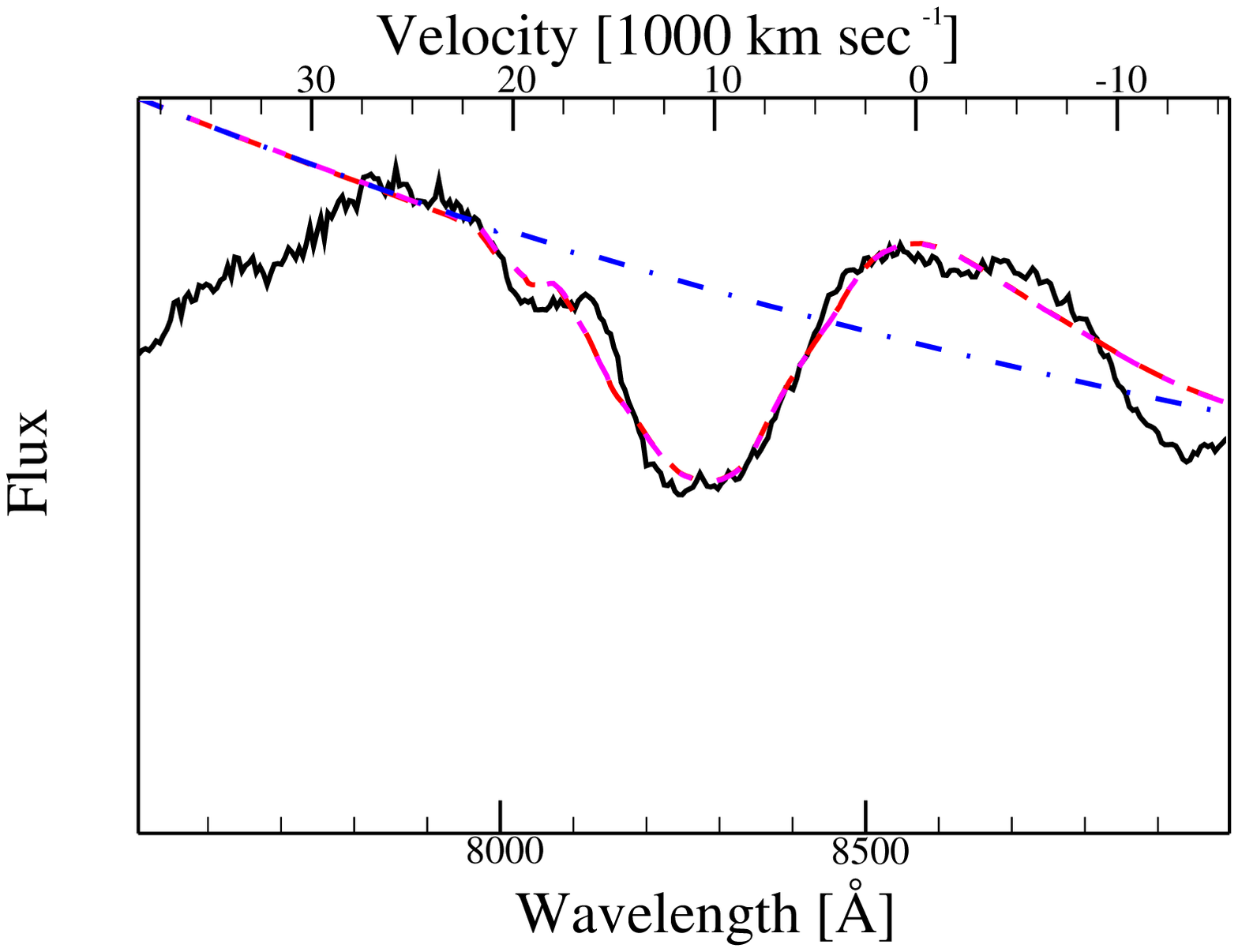}}\caption{\label{run57spectra}Evolution of the CaNIR feature in SN~2011fe comparing the observed data (black) and synthetic spectra generated from the model with a shell of 0.005\Msun (red). Figure labels indicate time relative to the explosion. The magenta long-dash--short-dash line shows the flux when the material in the shell is excluded, and the blue long-dash--dotted line shows the flux when the material in the ejecta is excluded. These demonstrate the relative contributions of the ejecta and shell components in each model. There is some effect due to the material in the ejecta at all epochs, but the material in the shell has a larger effect through the first $\sim$2.5~d. The material in the shell has little direct effect on the feature at 10.7~d and later. The absorption feature near $7500\Ang$ is presumed to be due to \ion{O}{I}; it was neither generated nor included in the fitting range. SN~2011fe data sources are listed in Table \ref{parametertable}. The velocity scale is based on the central component of the CaNIR triplet.}\end{figure*}

The temporal evolution of spectra is shown in Figure \ref{run57spectra} for the shell of mass 0.005\Msunp. These model spectra reveal that there is noticeable contribution from the shell through $\sim 10$~d. The model with mass of 0.005\Msun tends to reproduce the HVF at $8000\Ang$ better than the other models due to the location of the contact discontinuity, though the feature in the synthetic spectra tends to be weaker than the observed HVF. At intermediate epochs (2~d -- 7~d) the blue-ward side of the feature in the synthetic spectra is a poor match for the observed spectra. This may be due to P~Cygni emission from the neighboring feature near $7500\Ang$ contributing extra flux between $\sim7500$ -- $8000\Ang$. Alternatively, a shell that does not have uniform distribution of \ion{Ca}{II} could explain the discrepancy between the observed and synthetic features during these epochs. This may be the result of a different spatial distribution of calcium within the shell than we have assumed or by radiative effects (e.g. shadowing) resulting in more calcium in the lower states of the CaNIR transition at lower velocities.

\subsection{Feature velocity evolution}\label{ssec:velev}The velocity of the components of the CaNIR feature are often determined by fitting it with a series of Gaussians \citep[c.f.][]{2015ApJS..220...20Z}. The shape of the CaNIR feature, or any other absorption feature within a supernova spectrum, is dependent on the structure of the absorbing material. The shape is therefore complex and non-Gaussian. Fitting such a shape with multiple Gaussians can lead to falsely identifying components that do not physically exist. Additionally, the continuum is not obvious in the UV and visible range of supernova spectra due to the density of P~Cygni absorption and emission associated with those features. For the CaNIR feature, this can lead to use of the P~Cygni emission of the nearby $7500\Ang$ feature, as well as the P~Cygni emission from CaNIR itself, as the blue and red sides of the continuum, respectively. This can further affect the apparent shape of the CaNIR feature, leading to falsely identified components.

\begin{figure}
\centering
\includegraphics[width=0.37\textwidth,angle=-90]{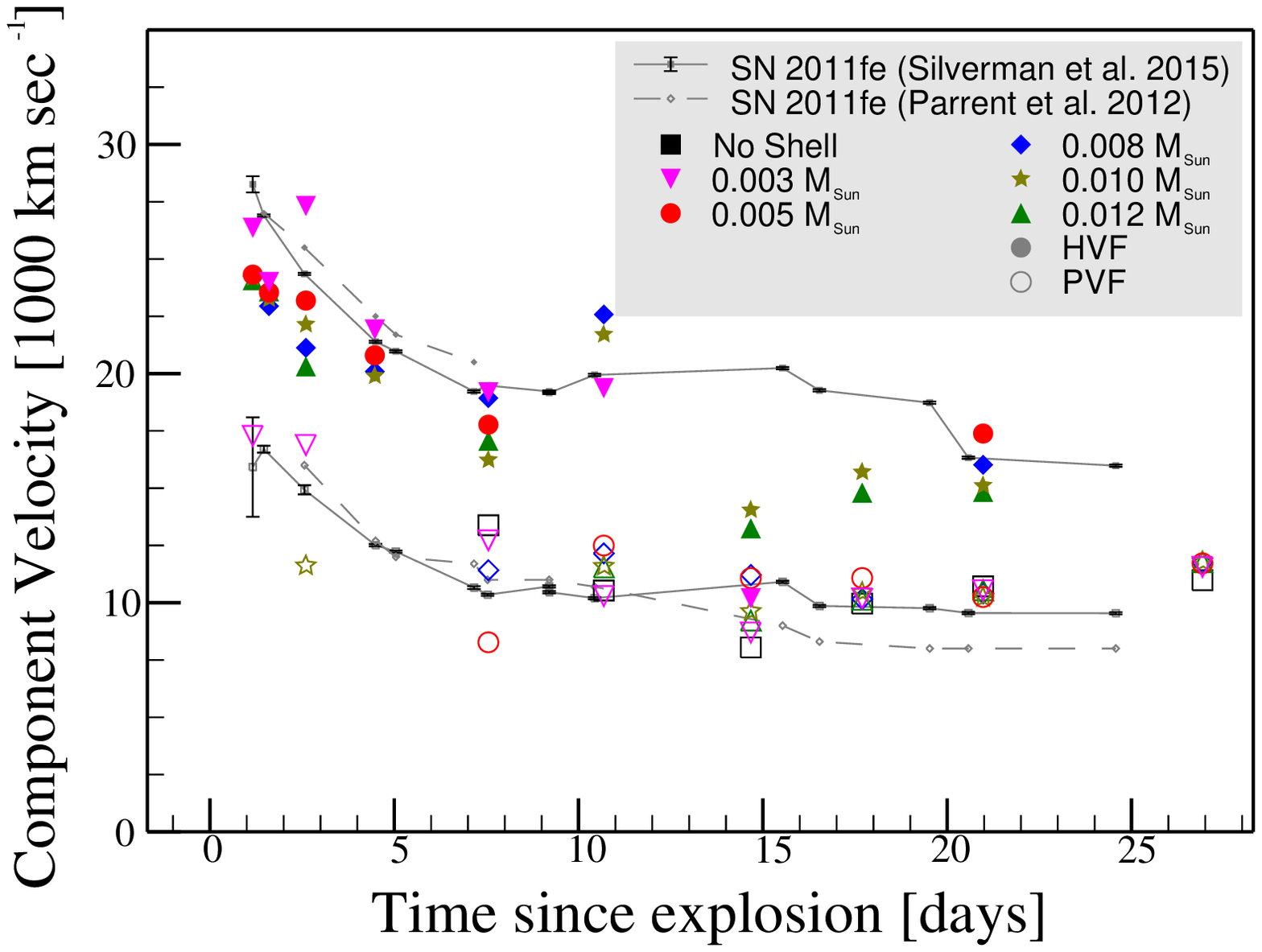}
\caption{\label{fig:velocityevolution}Evolution of the CaNIR feature velocity in SN~2011fe for each model based on Gaussian fitting, the minimum velocities $(v_{\mathrm{min}})$ for the PVF and HVF components of \citet{2012ApJ...752L..26P}, and the PVF and HVF component velocities reported in \citet{2015MNRAS.451.1973S}. Lines are provided to guide the eye to the observed velocities and highlight the PVF and HVF components. The uncertainties of the observed velocities reflect only instrumental error and do not account for uncertainty in the selection of the continuum. Two-component fits of the synthetic spectra occur at later epochs despite there being little to no absorption within the shell at these times. In these cases the apparent HVF are a result of the non-Gaussian shape of the overall feature.}
\end{figure}

Because we consider only absorption and related P~Cygni emission from \ion{Ca}{II} in this work, the synthetic spectra lack the $7500\Ang$ feature so we cannot exactly replicate the methods used on observed spectra. To emulate the observers' method, we use a continuum-flattened spectrum and identify the wavelength range over which absorption is greater than 1 per cent, thus avoiding the P~Cygni emission. We then perform a Gaussian fit using the method of \citet{2015ApJS..220...20Z}, allowing for either one or two sets of three Gaussians (three for the individual lines of the CaNIR feature, and one or two sets allowing for either a single component or PVF and HVF). Figure \ref{fig:velocityevolution} shows the result of these fits. If a single component results in a lower variance than two components, only one is shown. Single components with a velocity greater than $15000\kms$ are designated as HVF. We show the velocities reported in \citet{2015MNRAS.451.1973S} and the $v_{\mathrm{min}}$ values specified for fits with \texttt{SYANPPS} from \citet{2012ApJ...752L..26P} for comparison to the observed feature. The overall trend of all three are similar but there is scatter between the different methods and between models. It is particularly notable that the CaNIR feature generated from models of shells with higher mass (0.010\Msun and 0.012\Msunp) are best fit by two sets of Gaussians after $\sim13\:\mathrm{d}$, yet there is little to no contribution from these shells at those epochs. The additional Gaussian components at these epochs are due to the non-Gaussian shape of the feature, not to two physically distinct features.

\subsection{Alternative single degenerate models}\label{ssec:altmodel}
We used a Chandrasekhar mass, delayed-detonation supernova model for the work just summarized. An important category of SN~Ia explosion models involve edge-lit double detonations whereby the explosion is triggered in a sub-Chandrasekhar white dwarf by the detonation of a thin layer of helium on the outside of a carbon/oxygen core. The required mass of helium is estimated to range from $\sim 0.05$ -- 0.1\Msun \citep{2010A&A...514A..53F, 2011ApJ...734...38W}, although the mass of helium might be considerably less if the helium is enriched with carbon and oxygen \citep{2014ApJ...797...46S}. In this latter case, if the helium is burned to a mixture of calcium, silicon, or iron group elements, and all energy is released into the kinetic energy of the envelope, the envelope will expand at $10,000\pm 2,000\kms$. This is slow enough for the outer ejecta of the explosion to overtake the expanding envelope, leading to an interaction similar to that with the initially static shell considered here. While they need to be examined more quantitatively, such models might thus satisfy the properties of the shell needed to account for the HVF. 

Our models also constrain progenitor models that predict, or are consistent with, very little circumstellar medium. Models in this category could be spin-up / spin-down models in which mass transfer has long since ceased \citep{2011ApJ...738L...1D, 2011ApJ...730L..34J, 2012ApJ...759...56D} or models in which isolated white dwarfs explode by pycnonuclear reactions \citep{2015MNRAS.448.2100C}. In the absence of a circumstellar medium, these models would not produce HVF by the mechanism modeled here.

\section{Conclusion}\label{sec:conclusion}We fit the observed \ion{Ca}{II} near-infrared triplet (CaNIR) feature of SN~2011fe through 27~d after the explosion with synthetic spectra generated from models of \citet{2017MNRAS.467..778M} for Type Ia supernova that have interacted with compact shells with a mass between 0.003\Msun and 0.012\Msun and a model without a shell. The ejecta alone cannot explain the high-velocity features, and the model with a shell of mass 0.005\Msun performs better than other shells for SN~2011fe at more epochs, though the variation in the quality of the fit is small between models that include a shell. In shells with a mass less than about 0.008\Msunp, the photosphere lies within the shell only within the first minutes after the explosion and interaction, making such a shell difficult to detect photometrically.

The optical depth of neither the shell nor the ejecta follow the $t^{-2}$ trend that is expected from freely-expanding material with a constant ionization and excitation state. This suggests that the state of the calcium is not constant or that the distribution of calcium in the lower states of the \ion{Ca}{II} near-infrared transition within the ejecta and the shell of the supernova is not uniform or at least does not exactly match that of the models. 

The velocity evolution of the CaNIR feature determined using Gaussian fits of the synthetic spectra demonstrate that the evolution broadly matches the observed evolution, though this method artificially generates high-velocity components due to an attempt to fit an inherently non-Gaussian feature with multiple Gaussians. 

The composition and physical origins of the shells that we have considered remain to be determined. The presence of such a shell prior to the explosion is inconsistent with models that lack a circumstellar medium (CSM) or have an extended CSM such as a wind. Such a shell might be consistent with some explosion models that invoke a surface detonation of a helium envelope around the progenitor or other models that include accretion onto the progenitor, though the accretion disc itself would be much less massive than the shells we have invoked. It may be possible to determine the source of the shell if its composition can be determined, an important issue we intend to pursue in future work.

\appendix
\section{Optimizing the quality of fit in 4-D parameter space}\label{sec:optfit}
The fitting method that we employ --- a refined grid search using variance as the distinguishing parameter --- presents challenges in ensuring that the parameters that result in the best fit of the observed spectra are the most representative of the physical conditions. There are two problems that we encounter when fitting spectra in such a parameter space. The first is that variance (or $\chi$-square) can be a poor indicator of the quality of fit of a spectrum when the resulting fit is not exact. There may be minima or maxima that occur in the feature of interest that are washed out by differences between the model and observed spectra on a larger scale. For example, consider absorption by a doublet in a typical absorption line in a star. A fit to such a feature by a single Gaussian would manage to capture the relative broadness and average depth of the absorption features, but would fail to capture the presence of two minima. Higher order statistical moments or ``$\chi$-by-eye'' may help discern the failure of the fundamental model. We investigated use of moments up to the $6^{\mathrm{th}}$ order but did not find any improvement in the fit for our models --- the variance was the best predictor of the quality of the fit.

The second problem that we encounter is that we do not know the topography of the parameter space. Use of a grid search may lead to choice of a local minima as the best fit, rather than a nearby global minima that resides over a narrow, nearby ridge. Is is not computationally practical to sample the parameter space at high enough resolution to ensure that we have truly reached a global minimum. Furthermore, because the model is not expected to perfectly replicate the physical characteristics of the supernova, and because there is some degeneracy between the parameters (i.e. there is overlap in the absorption due to the shell and ejecta that can make them individually difficult to distinguish, and there is also degeneracy between the different models due to the effects of velocity of the photosphere and the relative strength of absorption by the shell or ejecta components) we expect the best fit within the available space to lie in a valley that is somewhat broad in those parameters that are degenerate. We do see this in the relatively small difference between the best quality of fit of the various models at a given epoch (c.f. Table \ref{tab:variance}). One of the ways to overcome this is through the choice of starting point of the search. By initially fitting by eye we can have some confidence that one point within our initial grid is likely to be within the deepest valley within the space.

\begin{figure*}
 \centering
 	\subfigure[Case `a']{
		\includegraphics[width=0.28\textwidth]{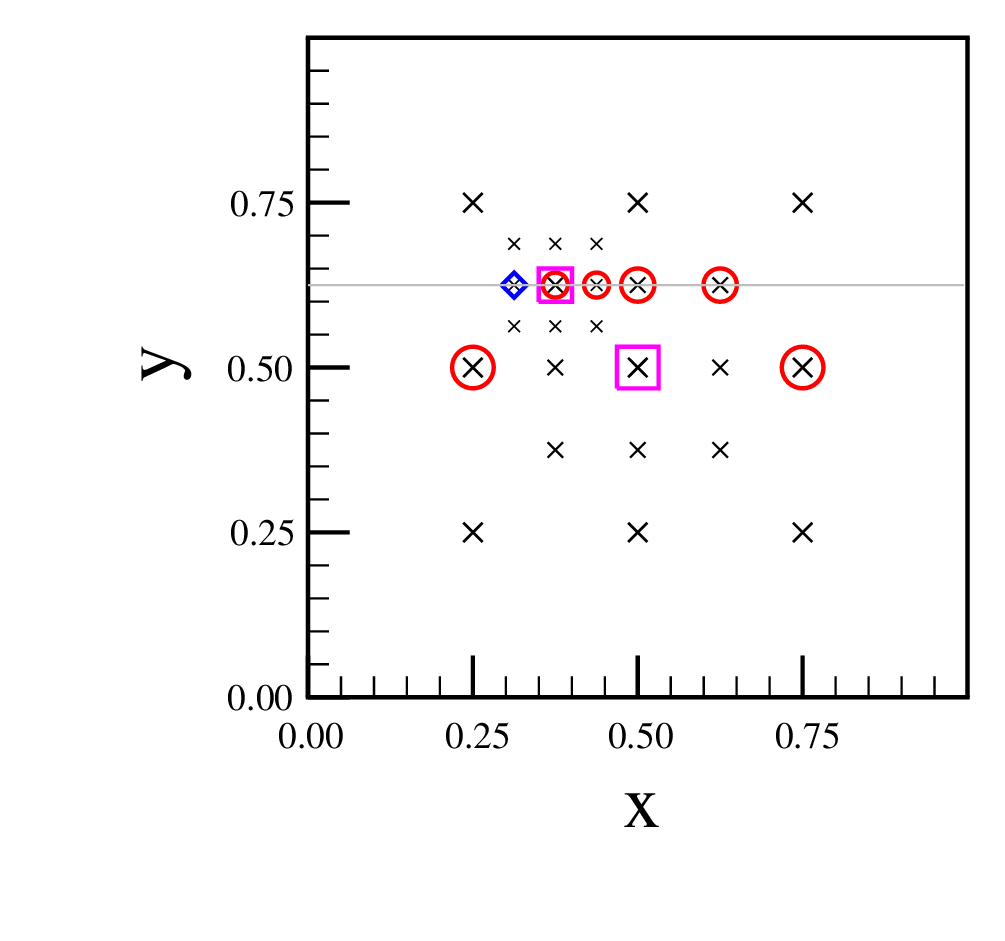}
		\label{FigA1:a}}
 	\subfigure[Case `b']{
		\includegraphics[width=0.28\textwidth]{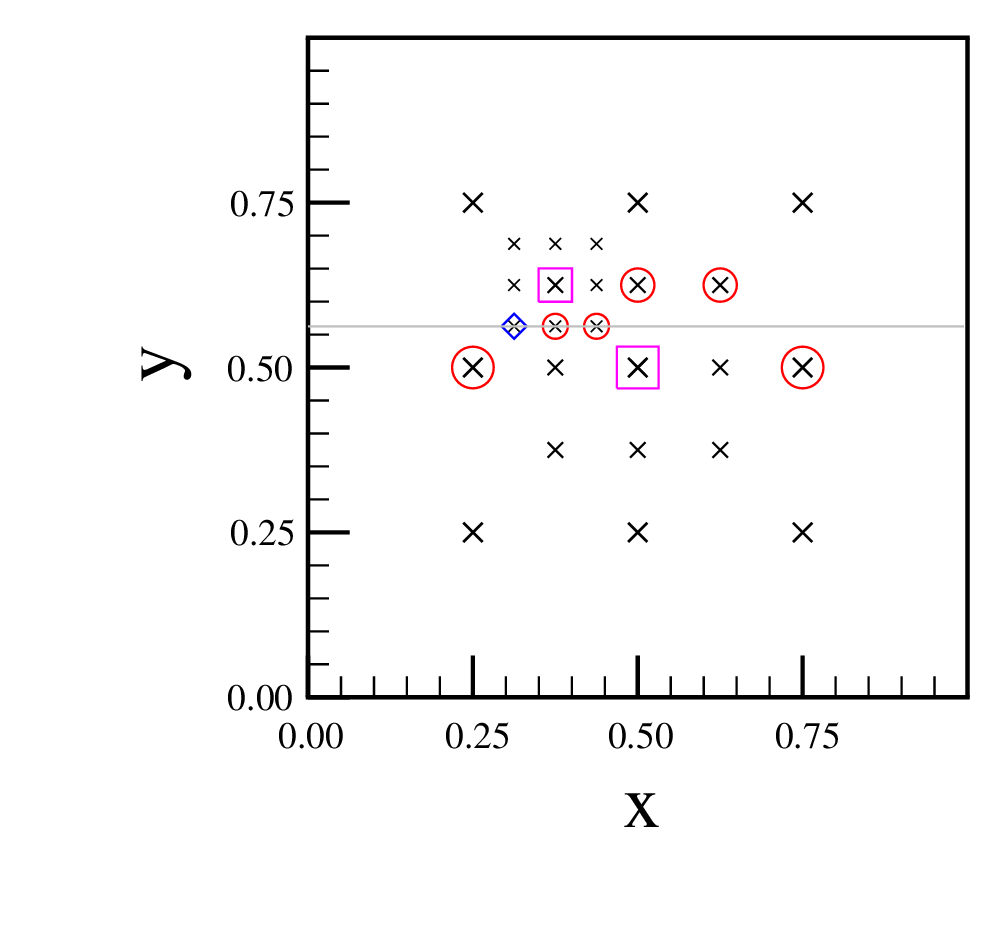}
		\label{FigA1:b}}
 	\subfigure[Case `c']{
		\includegraphics[width=0.28\textwidth]{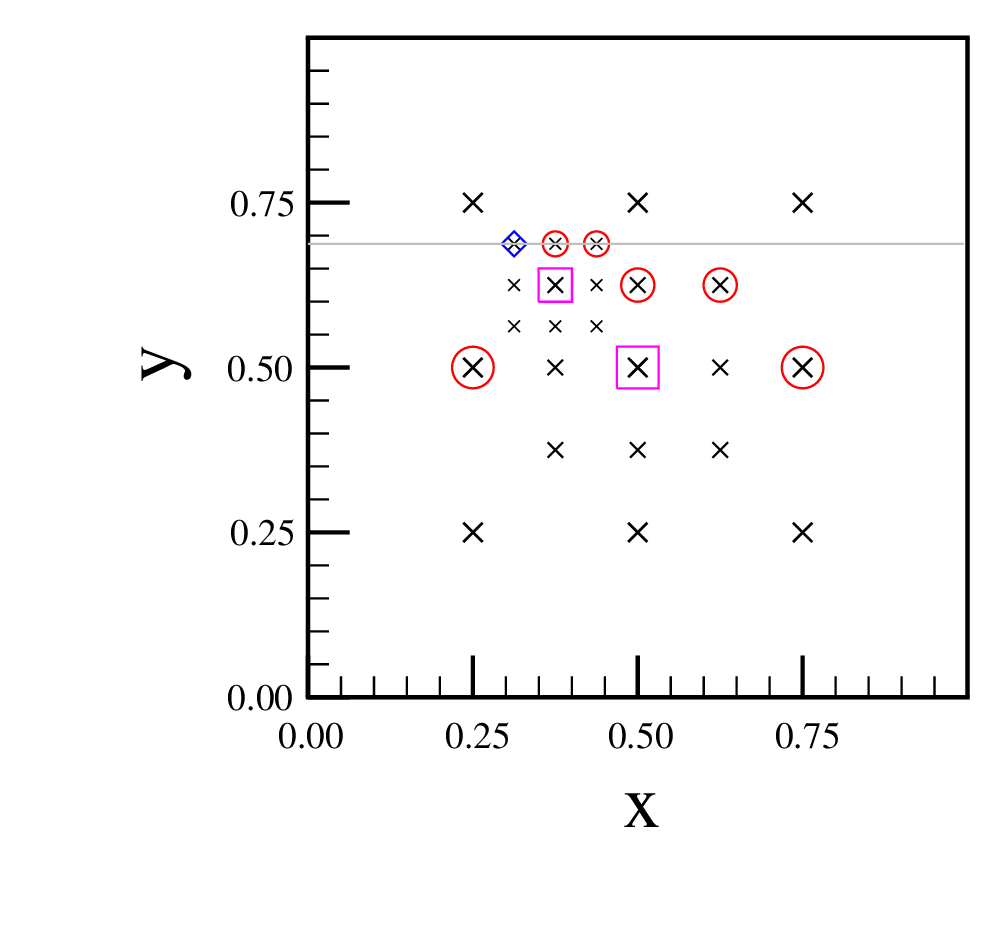}
		\label{FigA1:c}}
\caption{\label{FigA1} Three cases of sampling an arbitrary 2-D parameter space using our refined grid method. The locations of the samples are each shown as an `X', with the size of the symbol indicating the refinement level --- the most coarse samples are larger than the most refined samples. The best parameter sample is indicated with a blue diamond; the grey line indicates its y-coordinate. The magenta squares indicate the parameter samples that were the best at prior levels of refinement that are also the central point of the later level of refinement. The red circles indicate the points in each level of the grid that are co-linear (or, in a higher dimensional space, co-planar) with the best fitting samples at each level. The latter are also the points chosen for visualization of the quality of fit along the grey line. The three panels are identical except for the location of the parameter sample that is the best fit. Case 'a' maximizes the number of points that are co-linear along the horizontal axis that are used for the visualization. Case `b' demonstrates the case wherein the best fit is between the samples in the first and second level of refinement and it seems natural to use the samples from the middle row of the first level for demonstrating the quality of fit near $x = 0.25$ and $x = 0.75$. Case 'c' demonstrates the case of the best sample lying closer to the first level row of samples (the top row) that did not include the best sample at that level of refinement. In this case, we still use the middle row of samples at the lowest level of refinement for visualization of the quality of the fit because of the use of the central sample of the middle row of the first level for the subsequent refinement.}
\end{figure*}

To investigate the topography of the parameter space of an individual model presents its own challenge --- visualizing a non-uniformly sampled parameter space. The method that we use to refine the grid results in sub-grid refinement during each refinement step. For example, consider a 2-D parameter space with an evenly spaced grid of points for which we wish to visualize the quality of fit along only one axis. If the parameters at the central point of the grid fit the data the best, then the next refinement step would be a grid centered about that central point with half the size of the original grid. If one of the corners of the refined grid then fit the data best, the next refinement would be a grid with one-fourth the size of the original grid and centered on the corner of the second grid. Such a method of refinement is demonstrated in Figure \ref{FigA1}. In the figure, three cases are presented, all of which show the second level of refinement centered on the central parameter sample in the first level or refinement, and the third level or refinement centered on the sample at the top corner of the second level of refinement. The only difference in the three cases shown in Figure \ref{FigA1} are which parameter sample at the third level of refinement has the highest quality of fit. In case `a' (Fig. \ref{FigA1:a}) the best parameter sample is co-linear with the middle row of the grid at the second refined parameter level, in case `b' (Fig. \ref{FigA1:b}) the best parameter sample lies between the middle row of the first refined parameter level and the top row of the second refined parameter level, and in case `c' (Fig. \ref{FigA1:c}) the best parameter sample lies between the top rows of both the first and second refined parameter levels.

When attempting to visualize the quality of fit, we select a plane in the parameter space that includes the parameters that result in the best fit overall. Only a few other parameters will lie in a plane with this sample --- those in the same plane of the most refined grid and perhaps a subset of parameter samples in the previous level of refinement. Interpolation between points, especially at the lowest level of refinement (most coarse grid), will result in substantial loss of information, especially in a non-linear space as we expect this to be. Therefore we select the parameter samples in each level of refinement that are co-planar with the best sample in that level for use in visualizing the topography. 

\begin{figure}
\centering
\includegraphics[width=0.44\textwidth]{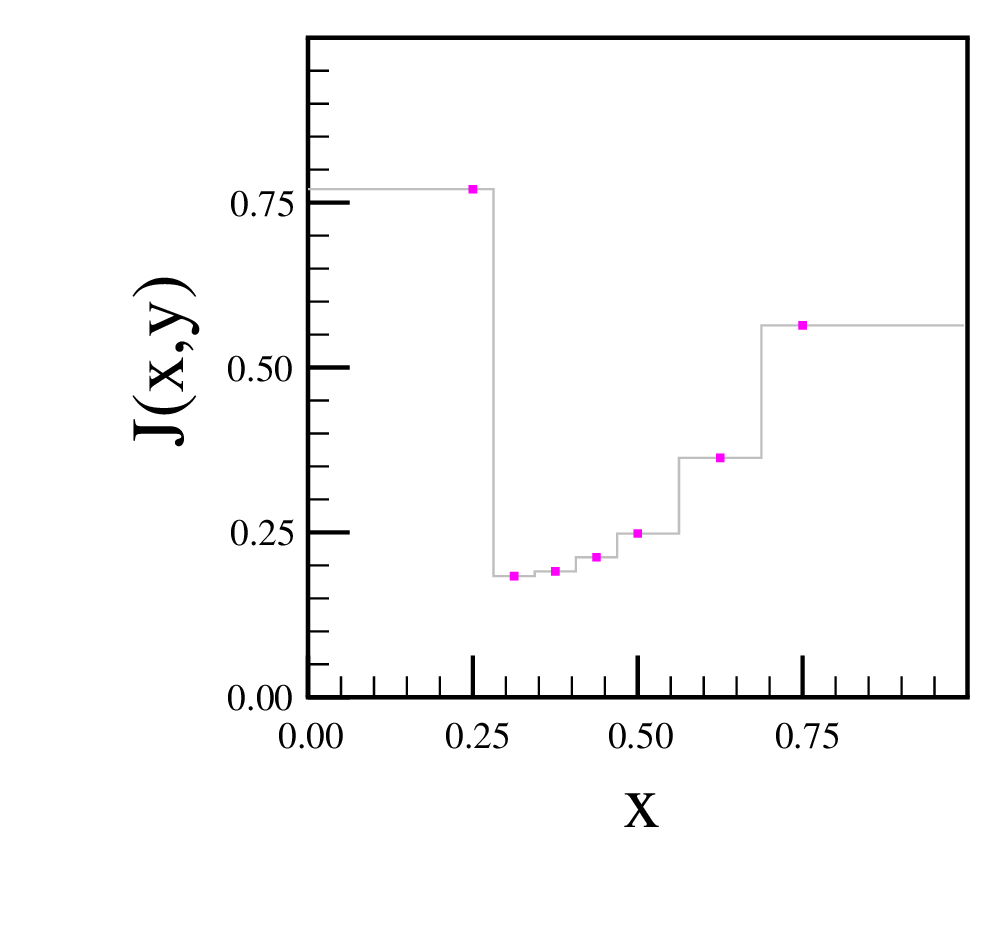}
\caption{\label{fig:exampleHist} A sample visualization of the quality of fit in the horizontal axis and sampled along the grey line for case `a' in figure \ref{FigA1}. The samples are taken at $x = 0.25,0.31,0.38,0.44,0.50,0.62,$ and $0.75$. Note that the samples are not co-linear in $y$ --- see text and Fig \ref{FigA1} for details. The segments are not centered at the $x$ values of the samples due to the non-uniform sampling; instead, the edges of the segments are halfway between samples. In this space, there is a steep dropoff somewhere between $x = 0.25$ and $x=0.31$. The best fit identified is at $x = 0.31$, though there may be a better fit between $x = 0.25$ and $x \simeq 0.42$. Additionally, if the quality of fit topography is very hilly, it is possible that samples are catching the crests of the hills and therefore a global minimum is missed. That the quality of fit monotonically decreases leftward from $x = 0.75$, the largest sampled value in this example, suggests that there are probably not additional minima; i.e. the probability of catching only crests in the samples between $x = 0.3$ and $x = 0.75$ is small.}
\end{figure}

\begin{figure}
\centering
\includegraphics[width=0.44\textwidth]{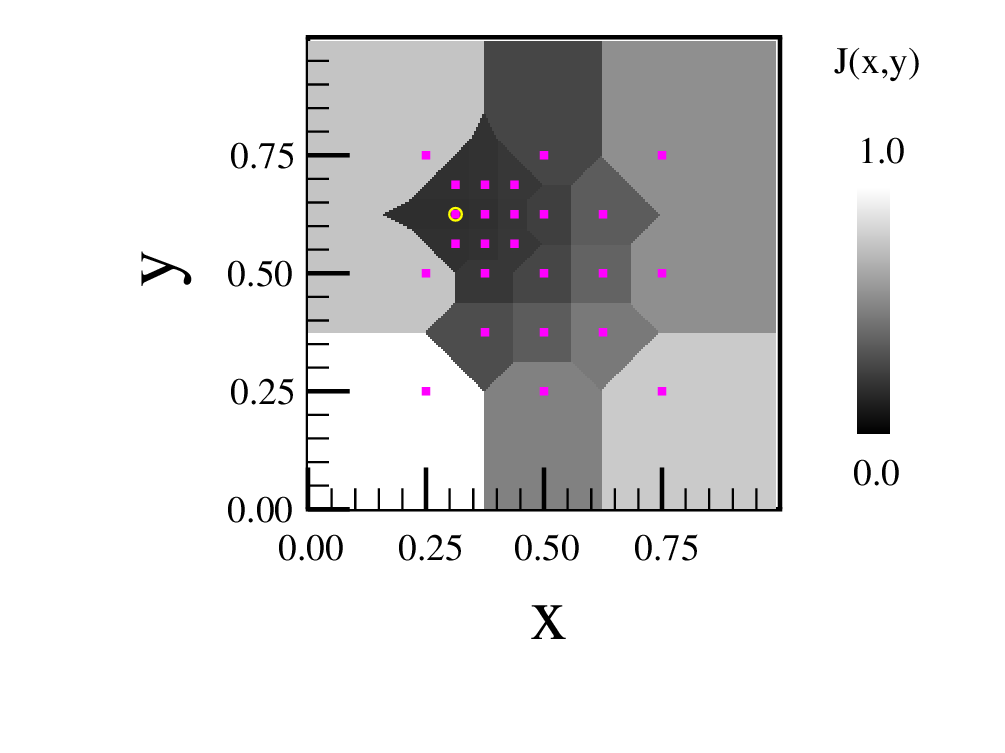}
\caption{\label{fig:example2D} A 2-D visualization of an example quality of fit of a parameter space that results in case `a' (See text and Fig. \ref{FigA1:a}). Here, shading is used to indicate quality of fit, with darker shades representing a better fit. The magenta points represent the locations of samples in the parameter space. The space in between samples is shaded based upon the nearest sample. The sample that resulted in the best fit is circled in yellow. This visualization suggests there is a bowl centered near the sample with the best fit and is in agreement with the 1-D visualization presented in Figure \ref{fig:exampleHist}.}
\end{figure}

Figure \ref{FigA1} also shows the parameter samples at each level of refinement that would be selected for visualization in this simple example of visualization along a line in a 2-D parameter space. In case `a' (Fig. \ref{FigA1:a}), there are eight unique parameter samples that are co-linear with the best sample (one sample is duplicated in two levels). Of these eight parameter samples, five of them are co-linear and one sample is ignored ($x = 0.5, y =0.5$), because there is a sample closer to or co-linear with the best sample. In cases `b' and 'c' (Fig. \ref{FigA1:b} and \ref{FigA1:c}, respectively), there are nine total parameter samples, though only seven would be used for visualization due to samples at higher levels of refinement being closer to the line on which the best sample lies. Of those seven, only three are co-linear with the best parameter sample. In case `c' (Fig. \ref{FigA1:c}) the best parameter sample lies between samples at the first and second level, but lies further from the sample that was the best of the first level. In this case, the middle row of the first level of refinement is still used for purposes of visualization because it included the best parameter sample at that level. Figure \ref{fig:exampleHist} shows an example of a 1-D visualization of the quality of fit of the parameter space for case `a', using the $x$ values of the selected samples, i.e. the samples circled in red in Figure \ref{FigA1}. Figure \ref{fig:example2D} shows a 2-D visualization of the quality of fit for the entire parameter space of case `a'. Both the 1-D and 2-D visualizations suggest that there is a smooth distribution in quality of fit with a minimum near the sample that result in the best quality of fit. 


\begin{figure}
\centering
\includegraphics[width=0.44\textwidth]{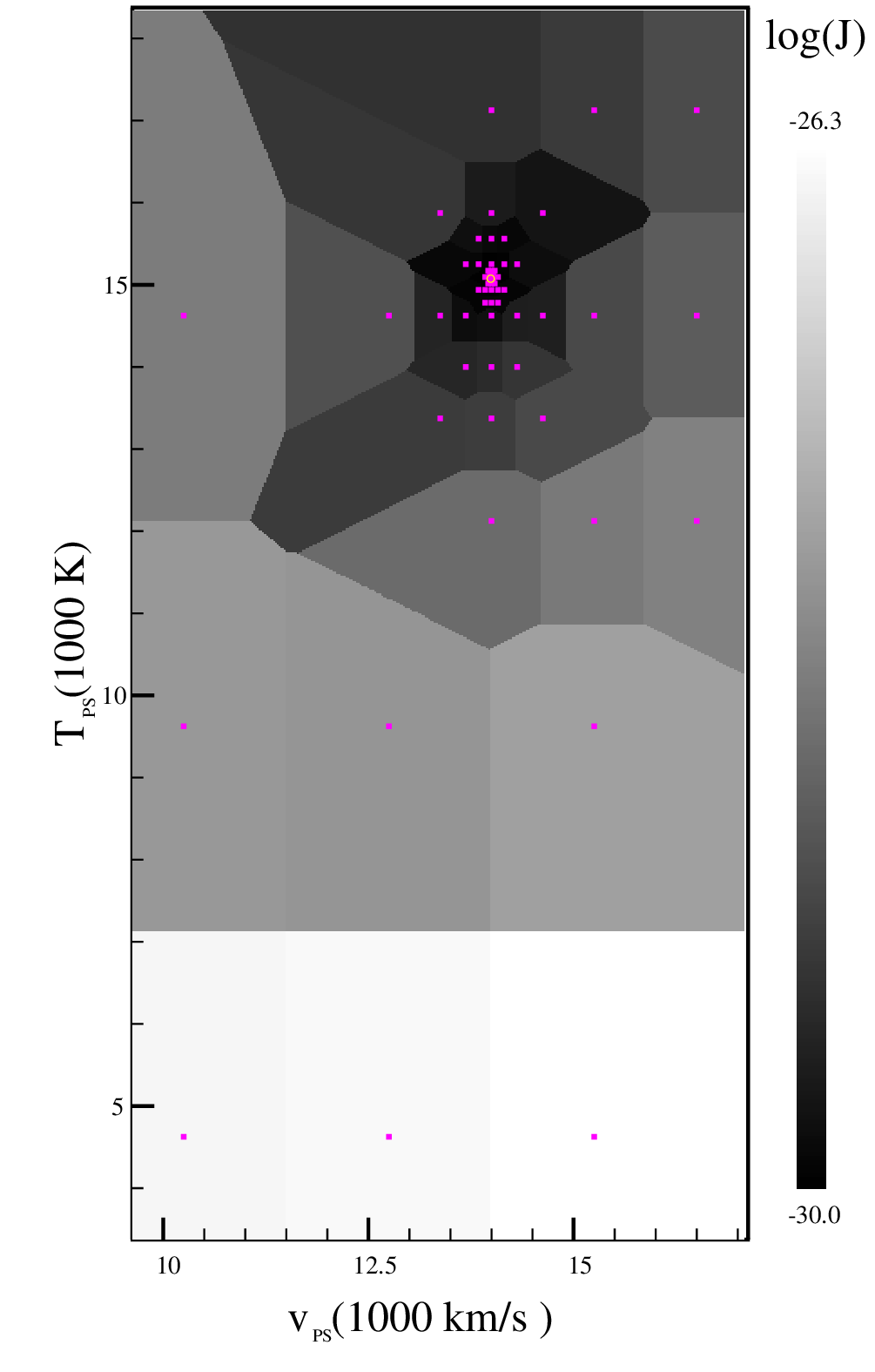}
\caption{\label{fig:fitTv}The quality of fit space for the model with a shell of mass 0.003\Msun at 7.55~d after the explosion in the photospheric temperature ($T_{\mathrm{PS}}$) and velocity ($v_{\mathrm{PS}}$) plane. Magenta points indicate the position in $(T_{\mathrm{PS}},v_{\mathrm{PS}})$ space of the samples, and the yellow circle indicates the sample that results in the best fit in the space. The shade indicates the variance ($J$) in units of $\mathrm{erg}^2\:\mathrm{cm}^{-4}\:\mathrm{sec}^{-2}$, with darker shades indicating better a quality of fit. The shading is in log scale, as shown on the plot, with the worst fits having a variance of $\log J \sim -26.3$ and the best having a variance of $\log J \sim -30$. Due to sparse sampling, the shading is based upon the nearest sample on the plane. In this space, there is a bowl near $T_{\mathrm{PS}}\sim 15,000\Kelvin$ and $v_{\mathrm{PS}}\sim 14,000\kms$. The first grid point selected for further refinement for this model is at the corner of the samples in the initial plane, with $T_{\mathrm{PS}}\sim 14,500\Kelvin$ and $v_{\mathrm{PS}}\sim 15,500\kms$.  }
\end{figure}

\begin{figure}
\centering
\includegraphics[width=0.44\textwidth]{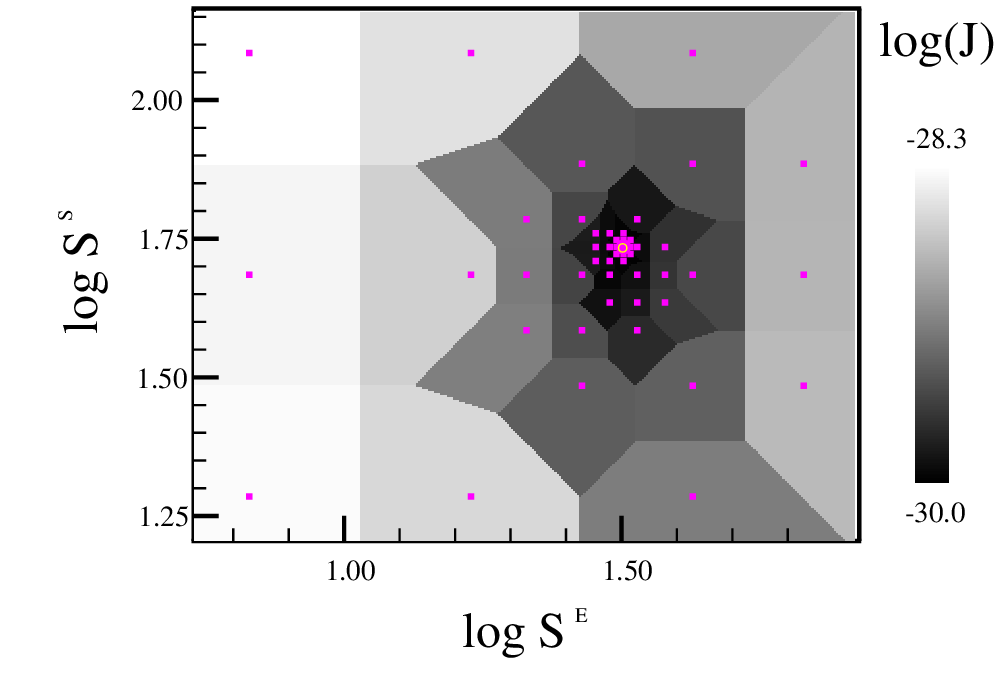}
\caption{\label{fig:fitES}A visualization of the space for the model with a shell of mass 0.003\Msun at 7.55~d after the explosion in the plane of the scalars for the ejecta ($S^\mathrm{E}$) and shell ($S^\mathrm{S}$). Magenta points indicate the position in $(S^\mathrm{E},S^\mathrm{S})$ space of the samples. The shade indicates the variance ($J$) in units of $\mathrm{erg}^2\:\mathrm{cm}^{-4}\:\mathrm{sec}^{-2}$, with darker shades indicating better a quality of fit. The shading is in log scale, as shown on the plot, with the worst fits having a variance of $\log J \sim -26.3$ and the best having a variance of $\log J \sim -30$. Due to sparse sampling, the shading is based upon the nearest sample on the plane. There is a bowl near $\log S^\mathrm{E} \sim 1.5$ and $\log S^\mathrm{S} \sim 1.75$. The bowl appears slightly elongated in the $S^\mathrm{S}$ axis.}
\end{figure}

In Figure \ref{fig:fitTv} we present the result of such a visualization for the model with a shell of mass 0.003\Msun at 4.47~d after the explosion for the temperature - velocity plane, where darker shades indicate better quality of fit. We find that there is a a bowl near $T_{\mathrm{PS}}\sim 15,000\Kelvin$ and $v_{\mathrm{PS}}\sim 14,000\kms$. The bowl appears relatively smooth, but it is possible that our result lies within a local, rather than global, minimum. Figure \ref{fig:fitES} shows the visualization of the quality of fit in the scalar parameter ($S^\mathrm{E}, S^\mathrm{S}$) axes. In these axes there is a bowl near $(\log S^\mathrm{E} \sim 1.5, \log S^\mathrm{S} \sim 1.75$. This bowl also appears smooth, and seems to be slightly elongated in the $S^\mathrm{S}$ axis.

Increasing the number of parameter samples taken at each level of grid refinement could increase the probability of finding a global minimum, though the computational workload scales as $N^4$, where $N$ is the number of samples taken in each axis. Using a suggested starting point for the grid requires use of an odd number of samples in each axis in order for the central point of the grid to lie at the starting point; therefore increasing the sampling in each axis from three points to five increases the number of samples from 81 per grid refinement level to 625 per refinement level, a factor of 7.7 larger. A grid with three samples in each axis requires approximately 240 CPU hours on Stampede2 at the Texas Advanced Computing Center to perform fitting at a single epoch for all six models that we have considered.

Overall, the method employed is a balance between computational effort and thorough sampling of the parameter space. The relative smoothness of the bowl in which the best samples lie suggest that the grid refinement method is reasonable for the space that we are sampling. Due to sparse sampling, however, uncertainties may still be larger than the spacing between points at the most-refined level of the grid. 


\section*{Acknowledgments}
This work was supported in part by NSF grant AST-1109801 and the Graduate Student Continuing Fellowship from the University of Texas at Austin. Thanks to the Texas Advanced Computing Center (TACC) at the University of Texas at Austin for providing HPC resources that have contributed to the research results reported within this paper. The software used in this work (\textsoft{SuShI/spectrafit}) is publicly available at \url{http://github.com/astrobit/snatk}.


\bibliographystyle{apj}

\bsp	
\label{lastpage}
\end{document}